# Hydrogen as a Renewable Energy Carrier for Commercial Aircraft

2.62: Advanced Fundamentals of Energy Transfer
Dr. Ahmed F Ghoniem

Caleb Amy and Alex Kunycky
May 12, 2019

# Contents

# Introduction

Aviation results in approximately 5% of climate change causing radiative forcing [1]. This is more than its share of GHG emissions (2.5%) because of the location of emissions, and other secondary effects. This effect is illustrated in Figure 1, where the key takeaway is that the total effect is about twice that of CO2 alone. Aviation emissions are not only a significant contributor to climate change [2], but they are also among the most difficult to mitigate, due the high performance requirements of aircraft, when compared to land and sea travel. Further exacerbating the problem is the rapid growth of air transportation, which is not expected to peak until 2075 with a projected $10^{13}$ passenger·kilometers traveled annually [3].

Most approaches to mitigate the consequences of aviation focus on the use of alternative liquid fuels, such a biofuels [4]. Instead, we focus on the use of hydrogen as an energy carrier. Potential benefits of hydrogen compared to synthetic liquid fuels include high mass based energy density, few processing steps, and potentially higher efficiency. The main downside is its inherently low volumetric density, due to it being the lightest molecule.

Hydrogen fuel has been used and demonstrated in multiple applications over the course of the last century, with varying degrees of success. From the Zeppelin airships of the 1930s [5], to the Tupolev Tu-155 hydrogen fuel airline demonstrator in the 1980s , to the Boeing Phantom Eye in 2013 [6], multiple aircraft have demonstrated the possibility of hydrogen-powered flight. The following is a review of these and other important technologies that could enable hydrogen-powered commercial flight.

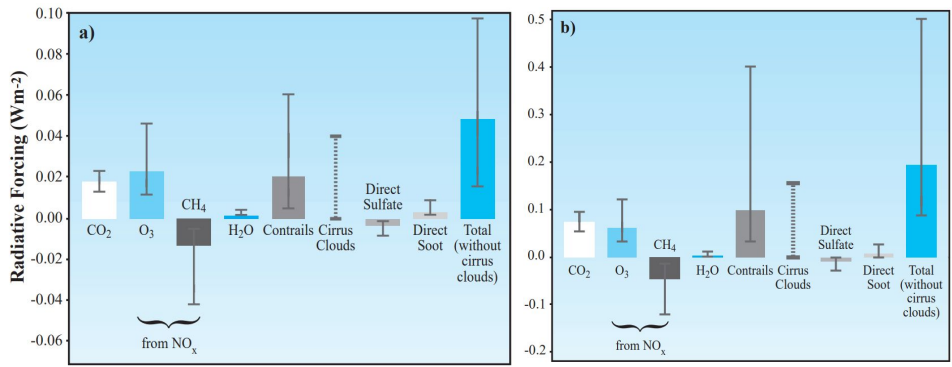

**Figure 1**: Aviation Radiative Forcing: a) 1992, b) 2050 projection [4]

# Production

Hydrogen is the most abundant element in our solar system [7] and has the highest mass based energy density of all chemical fuels [8]. Due to its availability and high ability to carry/store energy, hydrogen is regarded as an attractive energy carrier. Another benefit of hydrogen is that it does not release $CO_2$ during chemical reactions, unlike hydrocarbon fuels, and so if it can be harnessed, it may reduce $CO_2$ emissions. However, on Earth, natural hydrogen exists bonded to other elements, mainly as water and hydrocarbons. Therefore, in order to use hydrogen, it must be produced from a naturally occuring substance. There are many approaches to separating hydrogen from the elements it naturally bonds to (usually oxygen and carbon). In the following subsections, several approaches will be evaluated considering cost, practicality, and sustainability.

## Steam Reforming

Steam reforming of natural gas, oil, and coal account for 96% of global hydrogen production [9]. This has historically been the most cost effective method to produce hydrogen, but unfortunately results in carbon being converted to $CO_2$. Therefore, as long as this process is used, hydrogen is not a clean energy carrier. In this process, fuel (for example methane ($CH_4$)) reacts with steam to form CO and $H_2$, then a water-gas shift reaction is performed to convert CO + more $H_2O$ to $H_2$ + $CO_2$. This process has historically been cost effective compared to electrolysis because the source of electricity for electrolysis is also natural gas, and it is cheaper to directly convert than to intermediately create electricity. However, as renewables enter the grid with much lower cost (especially low variable cost) than natural gas [10], economics may eventually favor electrolysis, as demonstrated by the case study below.

## Direct Thermal Cracking of Methane

Another method of separating hydrogen from the carbon in methane is to simply heat methane. The key benefit of this approach is that the carbon can be captured as a dense solid (carbon black) instead of $CO_2$. In this way, methane cracking is regarded as a way to convert methane into a sustainable fuel--save for its limited supply. There are serious technical challenges to this approach, despite its seeming simplicity. First, the process occurs at very high temperatures, around 1200°C, and some of this heat should be recovered to maximize efficiency. Another challenge is that the carbon that separates tends to deposit and clog pipe walls in the methane heater/cracker.

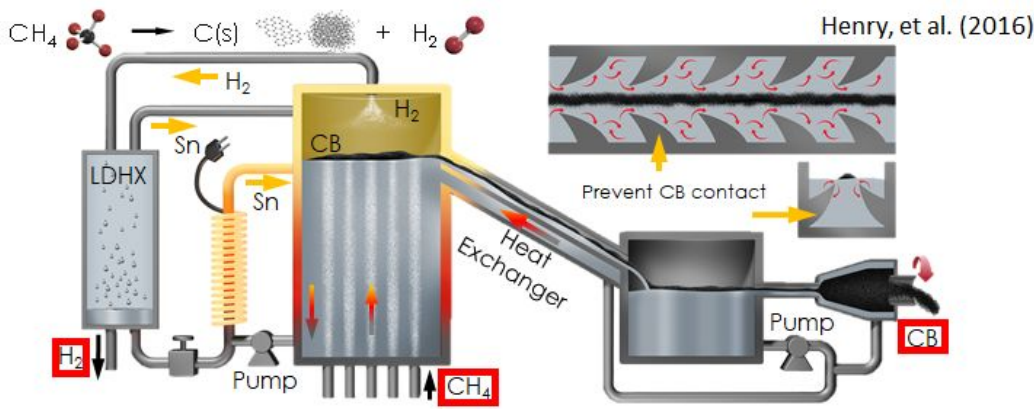

**Figure 2**: Methane Cracking concept using pumped liquid tin (Sn)

One way to resolve this clogging issue is to keep the carbon from reaching solid surfaces while it is hot. Wetzel has demonstrated this idea by bubbling methane through molten tin--which also results in very efficient heat transfer [11]. Henry has also proposed a new concept, shown in Figure 2, which incorporates pumping to aid in heat recovery and transport of carbon [12]. This technology could eventually replace methane reforming, especially if there is a strong market for the carbon produced (e.g. for manufacturing) or if a economic value is placed on clean energy. However, like methane reforming, it competes with ever cheaper electrolysis as wind and solar prices continue to fall.

## Direct Thermal Cracking of Water

The most obvious method of obtaining hydrogen is from water, and it is technically possible to do this simply by adding heat. However, unlike in methane cracking, where the temperatures required can be practically met, water does not appreciably split until about 2,500°C. Also, unlike in the case of methane cracking, water splitting necessarily results in free oxygen being released, so the infrastructure that was proposed for methane cracking would fail immediately from oxidation/burning.

Instead of directly thermally cracking water, current related research focuses on using thermochemical processes and catalysts to reduce this temperature drastically. These approaches typically involve redox cycles, (e.g., using ceria [13]), but have so far demonstrated low efficiency and energy density/reaction rates.

## Electrolysis

Electrochemical water splitting is a promising technology that has been widely demonstrated at efficiencies of 80% based on HHV (67% based on LHV, or ~180MJ/kg$H_2$). For large, deployed Alkaline Electrolyzer technology with high capacity factors (>50%), the majority of the cost of production is made up electricity, under typical US electricity prices [14][15]. As of 2011, capital expenditures (CapEx) and operating costs (OpEx) made up only ~20% of the cost of $H_2$ (total ~$3/kg) for a capacity factor of 97%. It is important to note, however, that if the capacity factor is low, for example 25% with solar photovoltaics (PV) as the sole source, the capital cost can dominate. This issue is explored in more detail in the LAX case study.

Given that the CapEx of electrolysis is significant, it is important to explore the state-of-the-art and expected future advancements. There are three main candidates: Alkaline Electrolysis Cells (AEC), Proton Exchange Membrane Electrolysis Cells (PEMEC) and Solid Oxide Electrolysis Cells (SOEC). There are also two main configurations of electrolyzers, each with its relative advantages. A unipolar design, shown in Figure 3a, resembles a large tank with the cathode and anode of many cells wired in parallel, resulting in high current and low voltage. Thus, in the unipolar configuration, modules are wired in series to increase the voltage. Bipolar electrolyzers, on the other hand and as shown in Figure 3b, have cells wired in series with bipoles between each. These are metal layers that act as both a cathode and anode with respective catalysts on each side. Unipolar configurations benefit from simple design and low maintenance whereas the bipolar design has higher voltage and thus less ohmic loss [16].

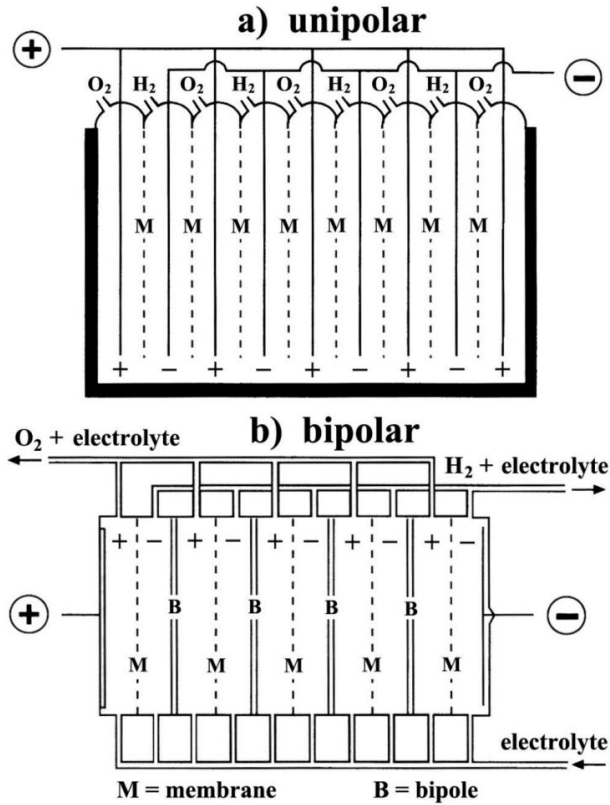

**Figure 3**: Electrolyzer stacks with (a) unipolar cell configuration where a single tank is used and the cells are in parallel and (b) bipolar cell configuration, where cells are in series and the bipolar metal sheets "B" are anode on one side and cathode on the other. [16]

AEC is currently the lowest CapEx electrolyzer technology, and the most widely deployed. These use a liquid (alkaline) electrolyte, potassium hydroxide (KOH) as shown in Figure 4. One major benefit of this technology, besides being the most mature technology, is that it doesn't require precious metals, unlike PEMEC. These systems also exhibit relatively long life, about 90,000 hours. AECs are operated near 80°C which helps to reduce the required voltage for water splitting and can operate at atmospheric pressure or ~35 bar.

PEMEC is a commercially deployed technology, although it is less mature than AEC. It relies on a solid, polymer electrolyte and can operate at higher pressure, current density, and efficiency than AEC. However, it requires precious metals (platinum, iridium) for catalysts, has shorter life, and requires higher purity water than AEC, which has so far made it less economical. There is significant ongoing work to improve PEMECs including thinner, stronger, longer lasting membranes that are composite (DSM -dimensionally stable - do not swell when wet), stamped metal bipolar plates, and modified catalysts which use 85% less platinum by alloying or thinner surfaces.

SOEC also uses a solid ion conduction, but relies on a ceramic instead of a polymer as the electrolyte and operates near 750°C. This higher temperature operation enables significantly lower electricity input (~⅓ less), which is related to the reduced bond strength of water at high temperature [17]. However, this reduced electricity is offset by the need for heat input, so ideally a waste steam would be available for this process. Historically SOECs have operated at low pressure, which reduces energy density but no precious metals are needed which could allow costs to be attractive once economies of scale are achieved. All things considered, it is not clear which of these electrolyzer types will ultimately be the most cost effective, and the decision may depend on the availability of waste heat, and the future cost of electricity compared to CapEx.

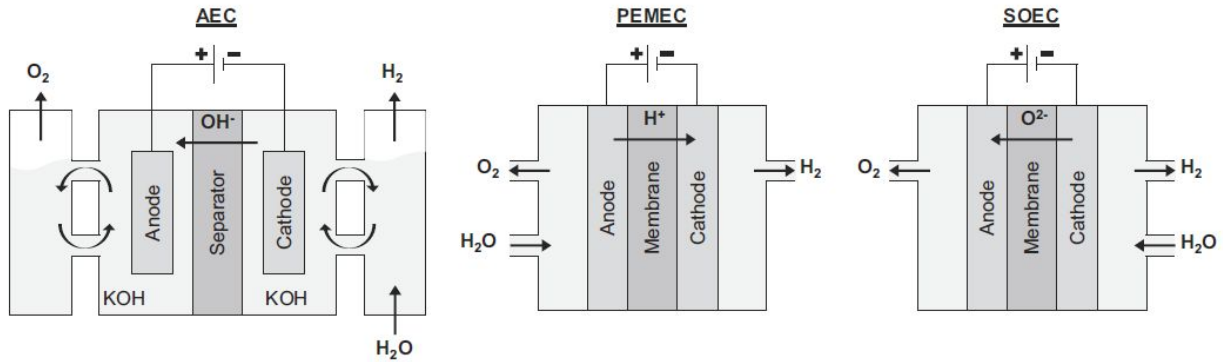

**Figure 4**: Conceptual diagrams of Alkaline Electrolysis Cell (AEC), Proton Exchange Membrane Electrolysis Cell (PEMEC) and Solid Oxide Electrolysis Cell (SOEC) [18]

It is also important to consider how electrolyzer voltage, and thus efficiency, change as a function of current density. As current density increases, the major voltage losses including electronic and protonic/ionic ohmic losses, mass transport losses, and activation losses also increase. As shown in Figure 5B, activation overpotential is a loss that must be accepted to achieve reasonable current density, then in the regime of operation ohmic losses dominate, followed by a steep increase in concentration losses which set an upper limit on current density [19].

Despite this, electrolyzers are often operated with significant losses in order to increase the power density and combat the high capital cost. It should be considered, however, that this decrease in efficiency causes an increase in electrical input, driving OpEx up. Therefore, a balance must be maintained between CapEx and Opex, or efficiency and power density. If advancements are made such that CapEx decreases dramatically, it will be ideal to operate at lower current density to reduce OpEx. On the other hand, if electricity prices fall dramatically, it will be advantageous to operate at higher current density to reduce CapEx. This concept is illustrated by Figure 5, where the end goal is to move toward the bottom right of the plot.

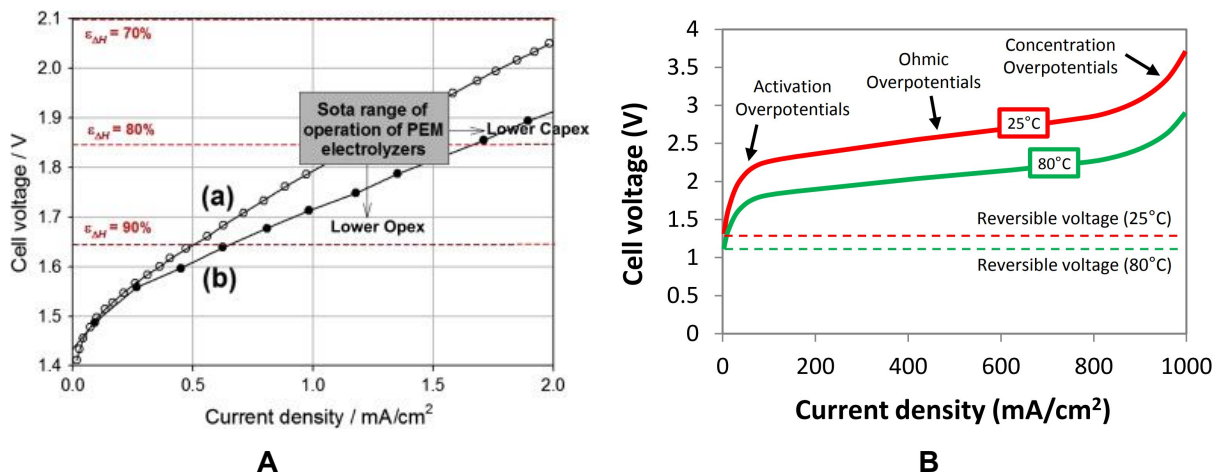

**Figure 5A**: The tradeoff between low CapEx and low OpEx - High current density decreases CapEx, but also decreased efficiency which increases electricity cost (OpEx) [20]. **B:** The major voltage/overpotential losses in the different regimes of current density [19]

As the cost of electricity from renewable sources decreases, even with relatively high capacity factors (e.g. 50%), the relative share of system costs will increase and will represent a price floor unless further advancements are made. We show below that these advancements (in addition to lower electricity prices) will likely be necessary for hydrogen to compete on with fossil energy (i.e. jet fuel), if emissions are not internalized. In this report, we take actual 2018 electricity tariff data [21] and current AEC costs/performance [22] as baseline inputs.

Nonetheless, we note that dramatic cost reductions in both CapEx and electricity prices appear to be on the horizon. For example, DOE 2030 electrolysis cost targets predict (based on demonstrated but not yet deployed improvements in technology, manufacturing, scale, etc.) a more than 50% reduction in electrolysis CapEx [23]. Similarly, recent solar and wind power purchase agreements (PPAs), which represent the levelized cost of *intermittent* electricity, have been as low as $0.025/kWh [10], which represents nearly a 50% decrease compared to our baseline costs. We predict that the stacking costs that have historically driven hydrogen far from competitiveness can be avoided in the current concept by locating hydrogen production on site at an airport and by eventually locating sustainable resources nearby in order to minimize transmission costs which account for an increasing share of electricity cost. These baseline costs alone (including the cost of capital but neglecting compression/liquefaction, storage, transport, etc) predict a possible hydrogen cost below $3.5/kg or nearly the same in gallons of gasoline (~Jet fuel) equivalent (gge) in terms of energy, before storage. A 2030 estimate considering reduce CapEx and electricity price could reduce this to $1.3/gge. For comparison, Jet fuel at a nominal cost of $2/gal. So hydrogen has a shot at competing directly with fossil fuels.

## Storage and Transportation

In the previous section, competing forms of hydrogen production were compared. However, to enable widespread use of hydrogen, the *total* cost of hydrogen must be compared conventional fossil fuels (e.g., gasoline). In order to make this comparison, the cost and losses associated with storage and transportation must also be taken into account.

Once generated, hydrogen must be stored until needed. However, in contrast to its high mass based energy density, hydrogen at atmospheric temperature and pressure has lower volumetric energy density than any conventional energy carrier. This is fundamentally because hydrogen is *the* lightest atom, consisting of just 1 proton, 1 electron, and no neutrons. For this reason it must be compressed or liquified for storage--which presents major cost and efficiency challenges due to the high pressure and/or low temperature required [24], as shown in Figure 6.

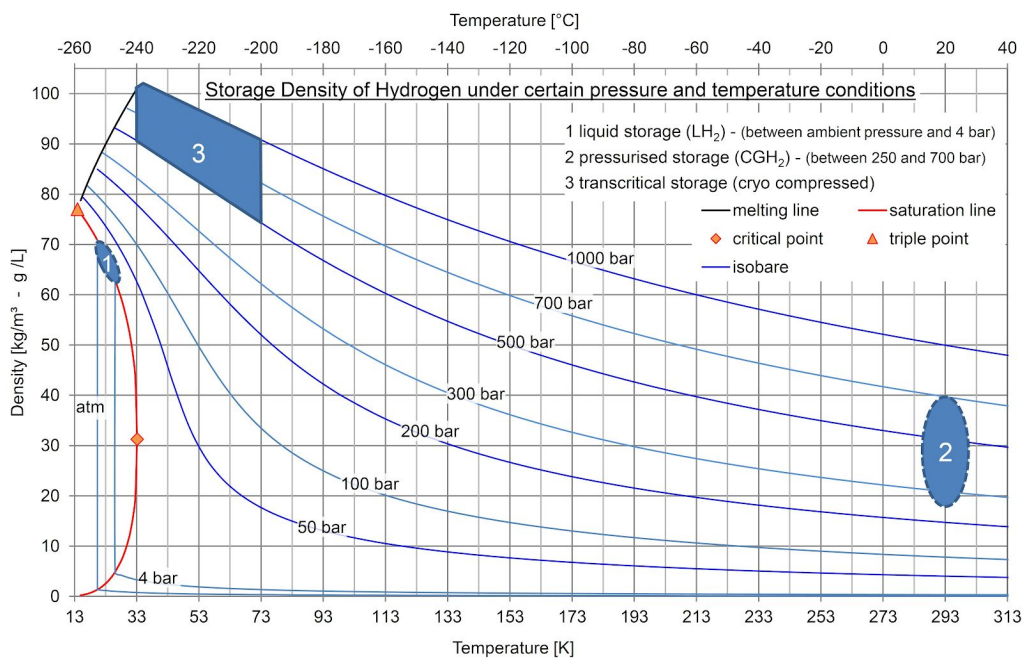

**Figure 6**: Storage Density of Hydrogen at various Pressures and Temperatures [25]

## Stationary Pressurized Storage

Hydrogen stored at atmospheric temperature is typically stored at a pressure of approximately 500 bar (region 2 in Figure 6). There are two consequences of increasing the density of hydrogen in this way. First of all, work is required to compress a gas, which in the ideal (isothermal) limit is defined as $W_{min} = R_u T \ln(P_2/P_1)$. In the case of hydrogen being compressed from 1 to 500 bar, a minimum of 8 MJ/kg is required to compress the gas. A more realistic estimate can be made by assuming a typical isentropic efficiency of 70% and a three stage adiabatic compressor. In this case, 15 MJ/kg is required. Notably, the higher heating value (HHV) of hydrogen is 142 MJ/kg, so this results in a 10% loss of availability. Assuming this compression work can be done by the baseline $0.053/kWh electricity source as the electrolyzer assumed above, this process raises the cost of $H_2$ (neglecting the cost of the pump) from $3.2/gge to $3.5/gge--which is likely tolerable.

The bigger issue comes with how to store this hydrogen, especially onboard an aircraft. Storage at these pressures is conventionally done in thick walled metal tanks, where the weight based storage efficiency is below 5% [26]. This can be improved moderately through the use of composites, such as composite overwrapped pressure vessels (COPV), which are projected to result in a storage cost of $600/kg [17]. For the intended application of storage at an airport, assuming weekly cycling, 30 year life, and a 10% discount rate, the tank increases the cost of $H_2$ from $3.5/gge to $3.7/gge. Notably daily cycling is unrealistic, especially if this system is expected to be resilient to cloudy/windless days. Weekly cycling would be more realistic, which increases the tank cost to $2.4/gge--above the cost of jet fuel on its own (although a direct comparison should consider the efficiencies of chemical-to-work conversion for each fuel). Clearly the tank cost is very important, and economies of scale do not help, since the material cost is essentially based on hoop stress, where tank volume scales with wall volume.

## Stationary Liquid Storage

Given that high pressure storage required heavy, expensive tanks an alternative method to consider is liquid storage. This resolves the need for heavy tanks, but adds the need for a thermally isolated tank and large energy costs to liquify. Liquid hydrogen is typically stored around region 1 in Figure 6. The minimum work to liquify hydrogen is $W_{min} = \Delta H_{fg} - T_o \Delta S_{fg}$, which is 12 MJ/kg This results in a minimum cost of liquefaction of $0.08/gge, although only as low as $0.27/gge has been practically realized, based on 30% efficiency [27]--bringing the cost of liquified $H_2$ to $3.8/gge. This can also be thought of as a 30% loss in availability, which is much higher than the pressurized case.

Of course, the CapEx of the liquefaction plant must also be considered. There is a wide array of possibilities for system types, but use of the Claude cycle, precooled by liquid $N_2$ is the most common today, which is an improvement on the Linde-Hampson Cycle. In the Linde-Hampson Cycle, hydrogen is compressed, sub-cooled, and throttled to generate Joule-Thompson cooling as shown in Figure 7. This process is typical of refrigeration cycles, although more extreme in this case because hydrogen has such a low boiling point (~30 K). Claude added a turbine to reduce the required work and to reduce or even eliminate the use of liquid $N_2$.

It is important to note that, unlike other liquefaction processes, hydrogen must also undergo a second transformation (in addition to physical phase change) from Ortho to Para state. The energy associated with this change is even greater than that of liquefaction, and happens relatively slowly. For this reason, catalysts are used within or between heat exchangers to accelerate the process. A version of this cycle is analyzed in the case study. The CapEx of these system is well understood due to significant deployment, at ~$90M/(kg/s). More detail is considered in the case study but, in short, Liquefaction capital cost is significant but currently overshadowed by electricity cost, especially because of low efficiency, therefore it is worth evaluating improvements to this process.

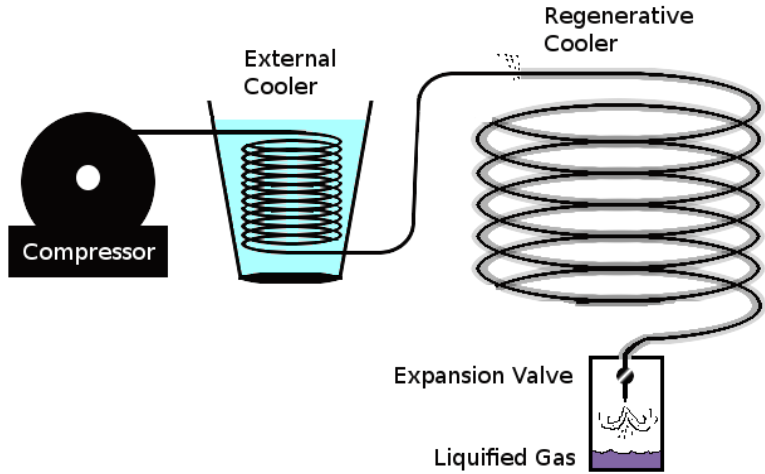

**Figure 7**: Simplest version of the Linde-Hampson Liquefaction cycle

Since heat transfer and insulation costs (which dominate in this low pressure, cryogenic temperature regime) scale with surface area, liquid $H_2$ tanks costs decrease as the size increases. Therefore, for a large installation (e.g. several days worth of hydrogen for a large airport), the tank cost actually becomes negligible [28]. For example, at LAX airport [29], about 100 20m diameter spherical tanks would be needed to store one week of $H_2$. The size and conditions of this tank is nearly identical to to those at rocket launch pads, such as for upcoming NASA SLS (previously Space Shuttle) in Florida [30]. These tanks operate at low pressure, and can be cost effectively vacuum insulated with glass bubble insulation. Recent research and experiments have shown that cost can be further reduced by eliminating boil off with active cooling by a closed helium refrigeration loop.[31]

## Transportation

Most transportation costs are avoided in the current concept of on-site generation, but considering some airports may not have the scale needed for cost effective production and liquefaction, the cost of transporting liquid $H_2$ is briefly considered. The most cost effective transportation method was found to be liquid $H_2$ on a truck, at about $2/gge for distances of a few hundred miles, on the same order as the entire cost of jet fuel--thus should be avoided if the goal is to compete with jet fuel [32].

Therefore, for the proposed use in airplanes, it is more effective, at least in the case of large airports, to generate hydrogen on site, thereby removing the need for long distance transportation, and enabling economies of scale in a single large storage tank.

Hydrogen still must be transported from a central reservoir to the planes, however. Due to the highly branched/distribution nature of airports, and the relatively low flow rate through each branch, which much be held at cryogenic temperatures, a branched pipeline is not ideal. Instead, this is done most cost-effectively using fueling trucks [33]. As the transport distance is short and the volumetric energy density of liquid hydrogen is about ⅓ that of jet fuel, local delivery is expected to cost about 3-5X as much as that of jet fuel, allowing for the increased cost of a cryogenic tank and multiple/larger fuel trucks [34]. This is explored in more detail in the following case study.

# Case Study of Hydrogen fueled LAX Airport

Converting the aviation industry from jet fuel to hydrogen would be a multi-decade, multi-trillion dollar process--but fuel cost savings and/or regulations could drive the transition. One way to get the process started may be to target a small subset of major airports so that planes which fly between them (eg. LAX, JFK, ATL, ORD) can prove out the technology. Another way would be for planes to store a round trip amount of fuel on board. This is feasible from a mass-based perspective, but challenging when the volume is considered. Therefore, in an admittedly non-conservative case, we investigate Los Angeles International Airport (LAX) as a potential site for sustainable hydrogen planes. We estimate the performance and cost of a sustainable airport enabled by hydrogen aircraft.

LAX is selected because it leads the world in fuel consumption [35]. As of 2012, more than 1.5 billion gallons of jet fuel was consumed at LAX, followed by JFK at 1.3 billion gallons. This corresponds to 45 billion kg $H_2$, on an energy basis. As we are investigating a self contained hydrogen infrastructure at LAX, we first consider the amount of hydrogen storage needed on site. Based on similar analysis [36], we conservatively estimate that 1 week of storage should be provided to balance out variation in prices, supply, or demand. Further, as discussed above, this study assumes on-site electrolysis generation, a central liquid $H_2$ storage tank with delivery to aircraft via fuel trucks.

## Production

### Electricity

For the base case, we apply the historical CapEx of the very mature alkaline electrolysis cell (AEC) technology of $1,000/kW, although PEMEC are approaching this CapEx [22]. We also take the typical efficiency of 80% based on HHV (67% based on LHV, or ~180MJ/kg$H_2$), as described above, based on a current density of ~600mA/cm$^2$. Similaring, in the base case, we consider the actual electricity tariff structure in place at LAX airport today. The detailed calculations are summarized below, and included in Appendix A.

LAX is located in the domain of the independent system operator (ISO) CAISO, which is responsible of scheduling and long-term planning of the electricity system. LAX specifically is served by the retailer Southern California Edison (SCE). SCE offers several tariff/rate options to large commercial consumers to help incentive them to consume electricity efficiency [37]. Although SCE does not (yet) offer real-time locational marginal prices (LMP, which is paid to producers) to consumers, they very roughly approximate it based on time of use (TOU) prices.

Here, there are different periods determined in advance when electricity will have certain rates. We select the TOU rate which charges a flat rate for 19/24 off-peak hours per day. During on-peak times, we do not consume at all. Therefore, the capacity factor of the plant is 19/24= ~79%, accounting for some time for maintenance as well. In 2019, the off-peak rate is $0.053/kWh, including $0.018/kWh for transmission/distribution [38]. There is typically a demand charge based on the MW required, but it can be avoided by participating in the base interruptible program (TOU-BIP). Electrolysis with on-site storage is a great candidate for this program because short (1-4 hour) interruptions, which are given a 15 minute notice, are no problem.

To consider the feasibility of local wind and solar generation, we briefly explore locating solar PV onsite and wind nearby. $H_2$ is consumed at a rate of 45 kg/s. Based on 80% efficient electrolysis and 30% efficient liquefaction, 10 GW is required to produce $H_2$ at this rate. Notably if this approach was pursued, the plant capacity factor would we reduce from 80% to 25-45%, which would cause a 2-3X increase in CapEx. This could only be justified with major (~50%) electricity savings.

### Solar

There is a fairly large amount of roof and grass space available at LAX that can be covered in solar panels. The airport itself covers 12 $km^2$ of land, and of that at least 20% is available for PV. Therefore, based on 2018 commercial solar module efficiency [10] of 19.1%, and average solar insolation [39] of 6.5 $kWh/m^2$/day this results in an average output of 125 MW. The high insolation of California is shown in Figure 8. However, LAX requires nearly 10 GW to meet fuel demands, so on site solar can't make a major impact.

### Wind

Onshore wind at an airport is impossible due to height regulations to prevent collisions with aircraft. However, nearby offshore wind speeds are very good. Also, offshore wind has a capacity factor [40] much higher than solar, around 0.45 in this area, within 20 miles of LAX, as shown in Figure 8. Unfortunately, the current cost of offshore wind, ~$0.09/kWh is too high to be competitive with solar, even considering the benefit of a higher capacity factor. Nonetheless, offshore wind costs are expected to decrease in the future, and will perhaps become even cheaper than onshore wind due to the possibility of larger blades (which can't be transported inland).

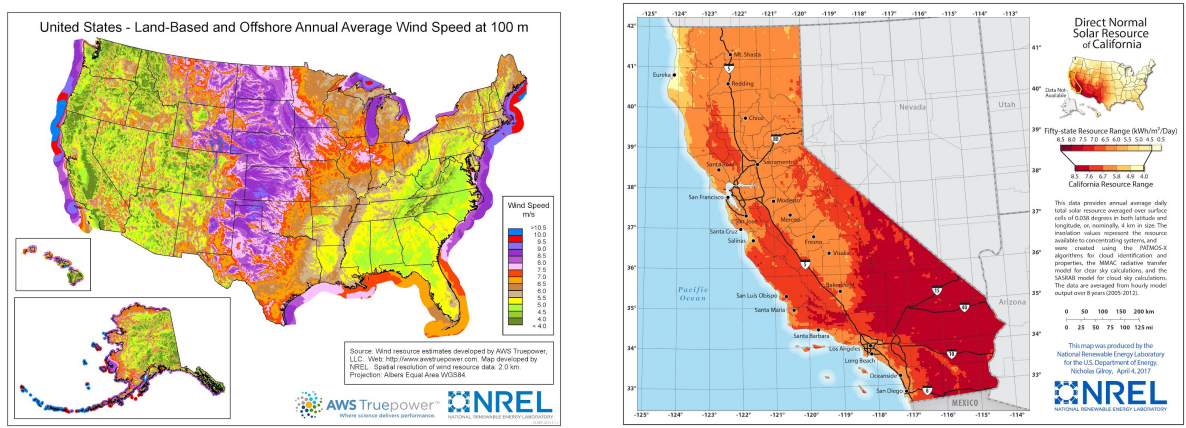

**Figure 8**: Wind and Solar Availability near LAX airport is among the best in the US

### Electrolysis

Based on the space constraints and current costs of wind and solar, we consider the actual available price of electricity on the grid as the input, which is currently $0.053/kWh based on the above analysis. Based on alkaline electrolysis (AEC) with an efficiency of 80% and

CapEx or $1000/kW, we estimate the cost of production. Electrolysis alone requires 8 GW average power. We estimate a 10% discount rate for the cost of capital and a useful life of 26 years. This includes 1 replacement of the cells at year 14, based on 90,000 hour cell life. This results in a CapEx of $0.78/gge. Given the average electricity load of 8GW and price of $0.053/kWh, the cost of electricity for electrolysis is $2.60/gge which is the *dominant* cost of the entire process. The relative costs of this base and future case are shown in Figure 14.

## Storage and Transportation

### Liquefaction

Since this hydrogen is for aircraft, liquid hydrogen will be used to avoid the very heavy tanks needed to contain high pressure gaseous hydrogen. Thus, the cost of liquefaction must be considered. Based on previous analysis of existing Claude+$LN_2$ systems by DOE [32], the CapEx is a concave down function such that specific costs decrease with scale. The rate of hydrogen required for LAX is far more than any demonstrated system or analysis, so we take a conservative cost based on a smaller (1/$20^{th}$ the size) plant, $90M/(kg/s). This, along with a 30% efficiency, sets the CapEx cost of liquefaction at $0.25/gge and the electricity cost (OpEx) at $0.58/gge. Overall, liquefaction raises the cost per energy from $3.3/gge to $4.1/gge.

These costs are quite significant, although they are only about ⅓ the cost of electrolysis, and the electricity cost will scale together (e.g. will decrease together if electricity become cheaper). It is important to note, however, than unlike in the case of electrolysis which is already 80+% efficient, there is much room for performance improvements for liquefaction. For this reason, we explore an alternative cycle in the future case following this section.

### Storage

To store 7 days worth of fuel on site at LAX airport [29], 93 20m diameter spherical tanks are needed. The size and conditions of this tank is nearly identical to to those at rocket launch pads, such as for upcoming NASA SLS (previously Space Shuttle) in Florida [30]. These tanks operate at low pressure, and can be cost effectively vacuum insulated with glass bubble insulation. Recent research and experiments have shown that cost can be further reduced by eliminating boil off with active cooling by a closed helium refrigeration loop. It has been shown that at scales as large as these, the tank cost actually becomes negligible compared to the rest of the components [28]. This amounts to $18/kg$H_2$ *stored* or $0.06/kg$H_2$ ($0.06/gge) on a levelized cost basis (assuming weekly cycling and 30 year life which has been demonstrated at NASA). Therefore, this increases the overall cost of $H_2$ less than 1%.

### Transportation

Long distance transportation is avoided by producing and storing hydrogen on site, which is a critical cost reduction as described above. The infrastructure required for refueling depends on the time available to refuel planes and how evenly distributed the refuelings are. Ideally refuelling occurs at distributed times, which is roughly accurate. Commercial liquid hydrogen trucks exist and hold 4300 kg [41]. Their cost is also known to be ~$700,000 per truck. Notably, although this is quite an expensive truck, the drivers actually cost more, assuming 2 drivers are needed per truck for 2 shifts and $50,000 salary.

Since the refueling time is only limited by the diameter of fuel lines, this time can be adapted to the required time, which is assumed to be 10 minutes. Based on major and minor fluid losses for liquid hydrogen [42], a hose of 2 inch diameter can provide this with only a 4 PSI pressure drop. For the truck to travel from the tank to the aircraft, 10 minutes allotted each way and 10 minutes for refueling the truck. There are 128 aircraft gates at LAX, and planes typically remain at the gate for 50+ minutes. With these conditions, we estimate that at least 20 trucks are needed on a perfectly distributed basis, but as many as 128 could be needed if all gates needed to be services simultaneously, since long-haul flights will require 2+ trucks. Therefore, as a conservative, but realistic estimate we plan to have 100 trucks and 200 full time drivers. In the end, this results in a negligible local transportation cost of only $0.01/gge.

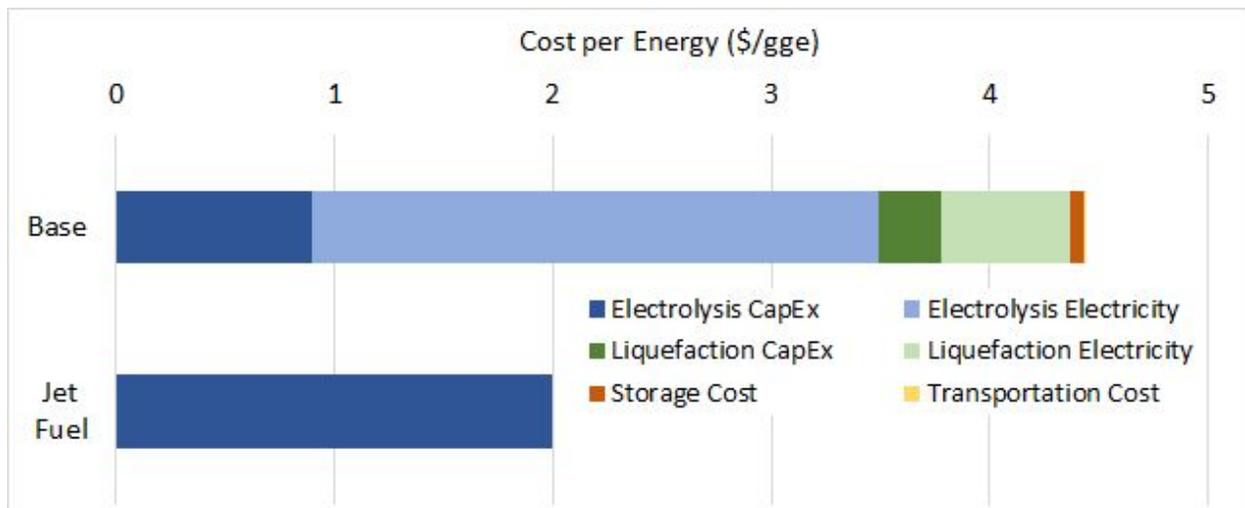

**Figure 9**: Baseline Hydrogen cost versus Jet Fuel. Hydrogen is more than twice the cost

## Technology and Policy Solutions for Cost-Competitive Hydrogen

Considering only deployed technology, current electricity tariffs, and realistic unsubsidized cost of capital, a "hydrogen LAX" is currently twice as expensive on an energy basis than jet fuel, as shown in Figure 9. The concept is not a dead end though, given the vast potential for improvements on political, economic, and technical fronts. In this section we explore these opportunities broady, then present a subset for further analysis to estimate the effect of their implementation.

### Political and Economic Solutions

A major capability of this concept is the ability to cheaply store energy for days and potentially even months. However, under the current regulations at LAX airport put in place by Southern California Edison, this capability is mostly wasted. Since the only time variable rates are predetermined time of use (TOU) rates, the plant can only really benefit from having 5 hours of storage to consume off peak. If instead electricity customers could pay actually locational marginal price (LMP) vast savings (and value to the grid) could be realized. The demand of this plant is nearly the same as the entire city of Los Angeles, but it is very flexible so it could help to balance demand. In fact, based on hourly 2018 LMP at LAX, if LMP was available to consumers, this system could have reduced electricity cost by more than 50%. This is true while maintaining a reasonable capacity factor (50%), and a reserve of 1 day of storage. Figure 10 shows the hydrogen storage volume over the course of the year, taking full advantage of storage by reducing usage when demand is high or supply (wind/solar) is low. The net result of this simple change to tariff structure would reduce the average electricity price from $0.053/kWh to $0.033/kWh, a major step toward cost competitiveness with jet fuel.

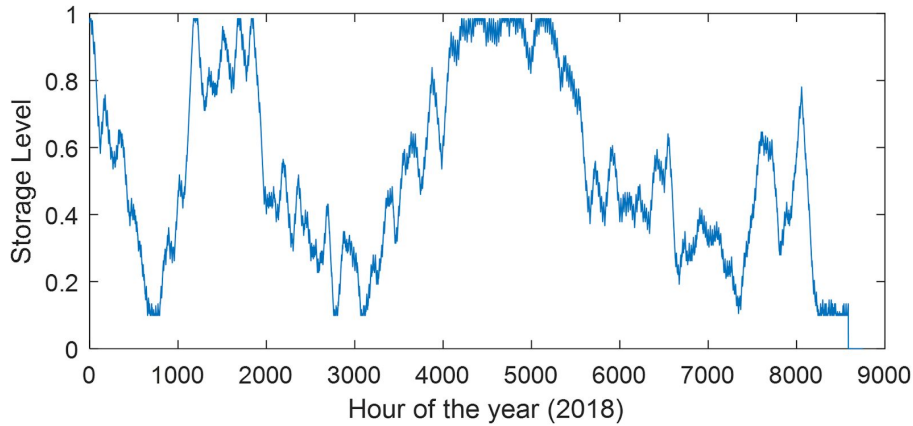

**Figure 10**: Relative hydrogen storage level throughout the year under optimum dispatch based on real locational marginal electricity prices

In the future, offshore wind may be located nearby LAX if it becomes a major load center. If this occurs, the transmission cost could be greatly reduced as the transmission lines would be very short. If transmission charges were reduced by 50%, which currently amount to $0.90/kgH_2$, the price of electricity would decrease even further to $0.024/kWh. It is important to note that capacity factor is important, but this drastic cost reduction in electricity would justify capacity factors as low as 20%. Furthermore, over time CapEx will decrease so buying off peak electricity will be more important. At this price, even without internalizing the social cost of carbon or other technology advancements the net fuel cost of hydrogen decreases from 2.1x to 1.3x the cost of jet fuel on an energy basis.

The above electricity savings assume no change in technology, but only pricing which better incentivizes customers to use truly off peak energy. Over the next decade, technological improvements will reduce costs. For example, offshore wind it 2-3x the cost of onshore wind today, but the costs are expected to decline by 30-50% as deployment grows in the US [43]. Offshore wind also has a relatively high capacity factor compared to solar and onshore wind, above 50% in some cases.

Another obvious, and logical, way for hydrogen to compete with jet fuel would be tax jet fuel for the negative externality of $CO_2$ emissions. The social cost of carbon is frequently estimate to (currently) ~$40/ton, which increases over time. If Jet fuel was taxed on this basis, its cost would increase by 20%, which along with the better pricing scheme nearly closes the gap with hydrogen.

## Technical Improvements to Hydrogen Technology

On the side of improving hydrogen related technology, there is also great opportunity for improvement. First of all, the costs used in the base case are *deployed* technologies, whereas recent advancement at a lower level of maturity are expected to generate significant CapEx reductions and modest efficiency improvement.

For example, the CapEx of electrolysis is expected to fall from $1000/kW to ~$400/Kw according to the NREL [17] and industry experts [18]. Much of these cost reductions are expected to come from manufacturing and supply chain improvements that occur naturally through economies of scale, especially for AECs considering we've already spent 100 years on the technology. However, there are also specific technical improvements that will enable these drastic cost reductions, especially for PEMECs. SOECs rely on more drastic technology leaps to become competitive, which are less predictable and so not covered in detail here. The major areas of improvement for PEMECs include cell design that can reduce CapEx and improve life and efficiency. First, the amount of precious metals must be reduced, and demonstrated approaches [22] exist for reductions of more than 85%. The challenge will be to use cheaper materials while also increasing their durability and efficiency, a challenge not dissimilar to the process lithium batteries are currently undergoing. Nonetheless, experts expect the efficiency of PEMEC to increase to ~85% while improving life to compete with AEC, ~90,000 hours.

The area with the more room for improvement from first order technology changes is liquefaction. Currently the efficiency of liquefaction systems is only about 30%, and the cost is quite high. As noted above, the Claude cycle improved on the Linde-hampson cycle, as shown in Figure 11, by the addition of a turbine/expander which helps to remove heat from the hydrogen and reduce the compressor work. This cycle includes a intercooled compressor, which is approximated as a isothermal compressor with 80% second law efficiency. The turbine is assumed to be 90% efficient and its exhaust is used to pre-cool incoming hydrogen. The heat exchanger is approximated as ideal and the fluid leaving the turbine is saturated vapor. States 2-4 are at 100 atm, while the rest are at 1 atm.

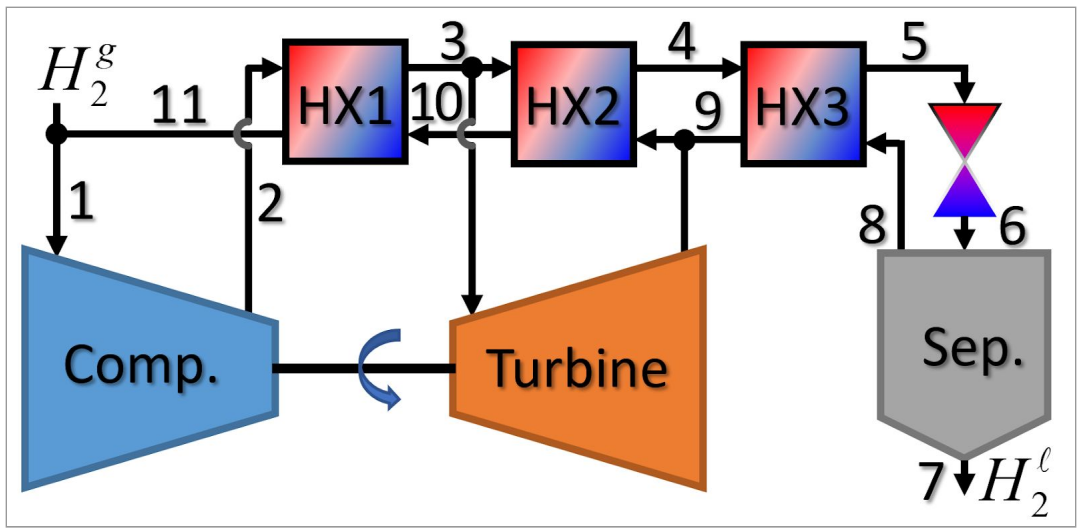

**Figure 11**: Simple Claude cycle, by adding a single turbine (4a) to a simple Linde-Hampson cycle. Figure adapted from [44].

| State | $T_i$ [K] | $P_i$ [kPa] | $h_i$ [kJ/kg] | $s_i$ [kJ/(kg*K)] |
|---|---|---|---|---|
| 1 | 300 | 101.3 | 3958 | 53.46 |
| 2 | 300 | 10130 | 4006 | 34.38 |
| 3 | 160 | 10130 | 2011 | 25.46 |
| 4 | 50.76 | 10130 | 466.1 | 8.881 |
| 5 | 39.98 | 10130 | 308.3 | 5.4 |
| 6 | 20.37 | 101.3 | 308.3 | 15.14 |
| 7 | 20.37 | 101.3 | 0.06483 | 0.003021 |
| 8 | 20.37 | 101.3 | 448.7 | 22.03 |
| 9 | 41.39 | 101.3 | 678.2 | 29.82 |
| 10 | 125 | 101.3 | 1594 | 41.79 |
| 11 | 300 | 101.3 | 3958 | 53.46 |

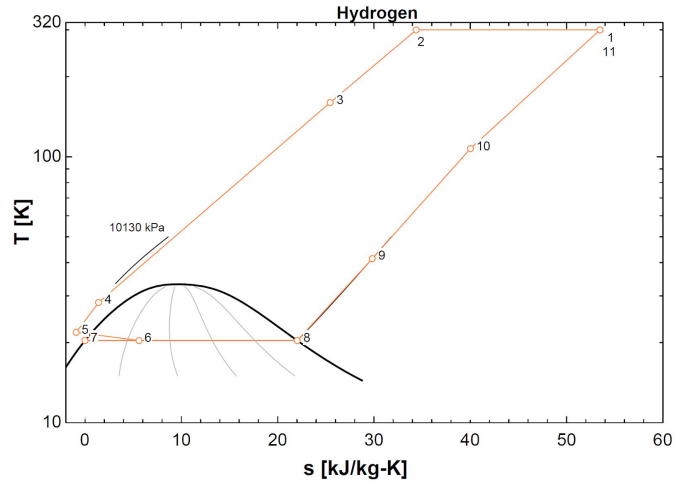

**Figure 12**: Thermodynamic states of a representative Claude cycle. Significant entropy is generated via the throttle valve, although not as much as in the Linde-Hampson cycle

The thermodynamic states are shown in Figure 12 Further, less heat exchange and compression is required, which helps offset the cost of the turbine. Although it was already known that commercial Claude systems achieve 30% efficiency, it is valuable to see where the entropy generation occurs, in order to address it. Many alternative thermodynamic cycles have been proposed throughout the 1900s [45] to increase efficiency, including the Kapitza, Heylandt, Pre-cooled Claude, Helium-refrigerated, and others but none have been commercialized for Hydrogen. Nonetheless, as the scale of hydrogen production increases, economies of scale may enable CapEx reductions that cause cost to be completely dominated by electricity cost, making efficiency improvements more important and affordable.

A further improvement, at least from an efficiency standpoint, would be to remove the throttling process ("isenthalpic" process) altogether. In this way, the system can most closely resemble a reversible/isentropic system, since each component has the potential to be reversible[45], [46]. A cycle which achieves this by cryogenically cooling helium without liquefaction and heat exchanging with hydrogen was proposed by PI Shimko and the MIT Cryogenics Lab in 2006. While the project continued until 2011 with promising results, the current state of this approach is unclear. In any case the proposed cycle was expected to have cost similar or slightly less than Claude liquefaction while increasing efficiency from 30% to 45%. A schematic of the Shimko cycle is presented in Figure 13. The simplest embodiment of this process would have just one heat exchanger, but due to the ortho-to-para conversion, it is preferred to proceed stepwise. In this way, the hydrogen is cooled, then is allowed to approach ortho/para equilibrium in each of the four catalyst chambers, which causes it to warm. Less entropy is generated if the hydrogen is cooled incrementally after cooling since this approximates isothermal cooling which would be cost prohibitive.

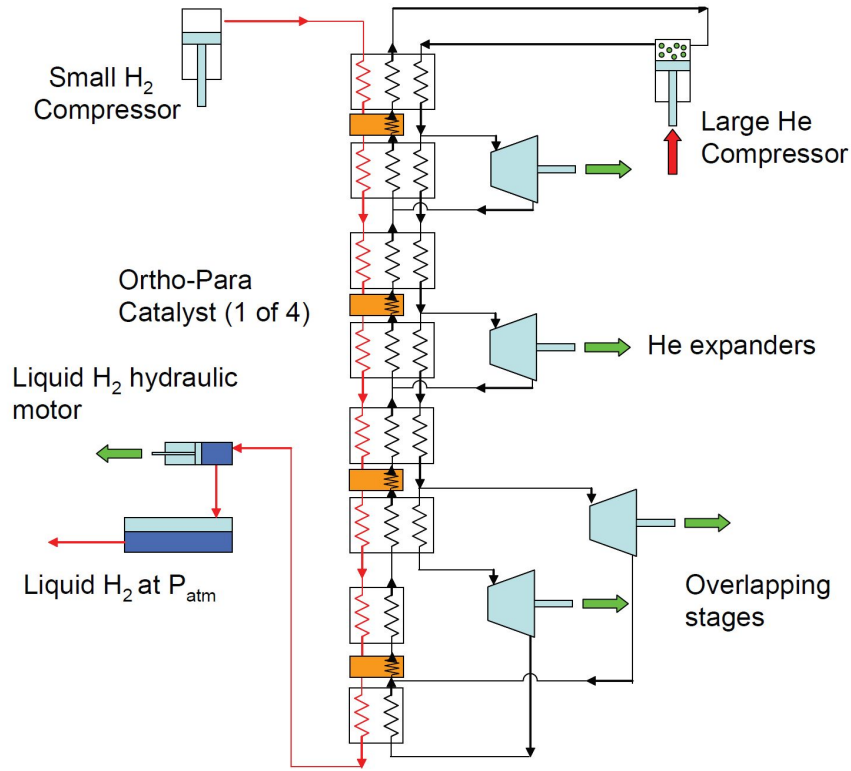

**Figure 13**: Throttle-less liquefaction of hydrogen by helium refrigeration [46]

One last, but very powerful, region for improvement is reducing the energy requirement of aircraft. Hydrogen is poised to enable this, especially if fuel cells are used to generate thrust on board the aircraft. As explored in the next section, the fuel requirements could be reduced by as much as 12%, which could nearly bridge the cost gap even without any of the above improvements.

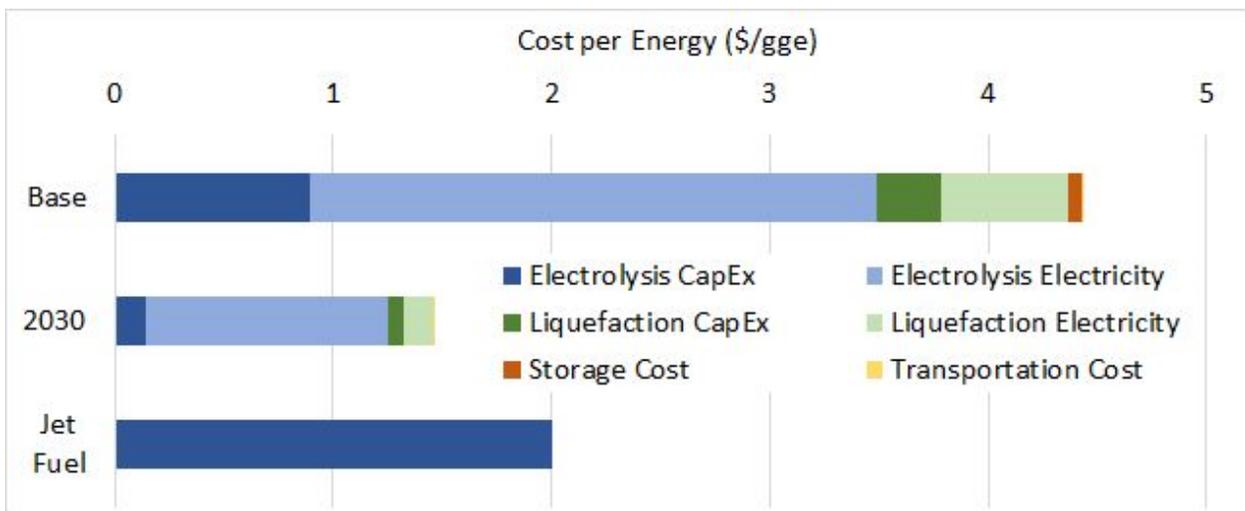

**Figure 14**: Cost of aviation fuel on an energy basis. Hydrogen today is more expensive, but could become cheaper than jet fuel over the next decade under plausible assumptions

# Hydrogen-Powered Aircraft Design

## Storage Tank Design

The cold temperatures required to keep liquid hydrogen fuel from boiling off require special equipment for fuel transfer and storage. The distribution of hydrogen fuel has been well studied for industrial and automobile applications [34] and explored here in the previous sections. Hydrogen storage onboard aircraft has also been studied, though with less detail and fewer applications.

One significant challenge of hydrogen storage is its volumetric density. Though liquid $H_2$ is around 2.8 times the mass based energy density of traditional kerosene jet fuels, it takes up about 4 times the volume [47]. This is one main reason that hydrogen-powered aircraft must have fundamentally different architecture than traditional kerosene-burning aircraft: the aircraft wings do not contain enough volume to serve the usual dual-purpose of producing lift and storing fuel, so the liquid $H_2$ must be kept in separate dedicated tanks. As a secondary effect, this would allow long-range hydrogen powered aircraft to utilize smaller wings with higher wing loading than current intercontinental aircraft [48].

As noted previously, heat transfer and material costs both scale with surface area. Cylindrical storage tanks would fit the profile of current aircraft fuselages and would maximize utilization of the available storage volume. If cylindrical tanks can be fitted with aerodynamic caps on each end, they can be carried outside of the fuselage entirely [47], not unlike auxiliary fuel tanks installed on many aircraft today. However, they have larger surface area for the same volume, increasing tank material weight significantly when compared to spherical tanks. In addition, a large single tank would still need inner partitions to keep the liquid fuel from sloshing during aircraft maneuvers [49]. Storage in a few large tanks likely offers the best combination of low weight and stable fuel storage while minimizing liquid $H_2$ boil-off.

Evaporation of liquid $H_2$ is another significant challenge of $H_2$ storage, especially for long-haul international flights. Liquid $H_2$ is typically stored around 20K, requiring significant insulation to maintain the fuel in liquid form with a manageable rate of boil off. Spherical tanks have been designed experimentally for liquid $H_2$ storage on aircraft with an evaporation rate of only 0.05% per hour, using a combination of vacuum layers, minimal point-contact supports, and gas/liquid transfer coils to significantly reduce the rate of heat transfer to the tank, as shown in Figure 15. For the purposes of this project, we assume that such tanks could be installed in FAA-certified aircraft and achieve comparable boil rates in production designs. The boil off rates change slightly based on tank geometry as shown in the later case studies, though overall boil rates as a percentage of total onboard fuel mass are fairly constant and, more importantly, well below fuel consumption rates even at idle/taxi power settings. Preliminary studies show that liquid $H_2$ can be stored efficiently, with hydrogen comprising up to 70% of the total (fuel + tank) weight in small regional aircraft, and nearly 80% for larger long-range aircraft [50]. Additional optimization may be possible considering other materials, variances in tank ullage and venting pressure, number of tanks used, and optimization for exact aircraft size.

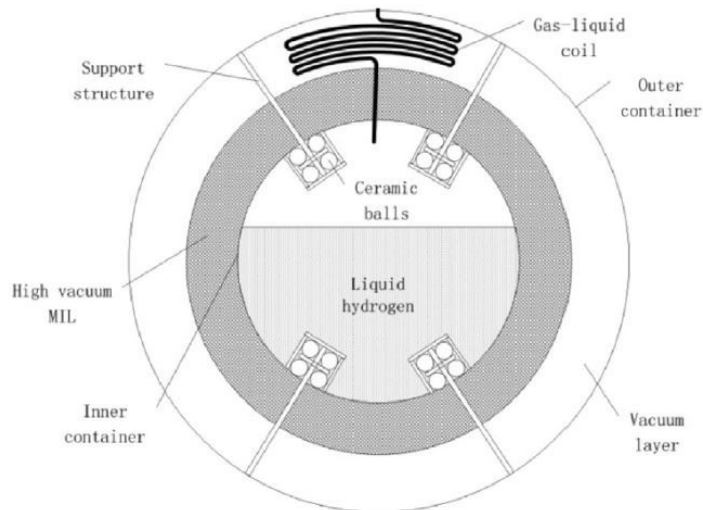

**Figure 15** Schematic of a spherical liquid $H_2$ storage tank with 0.05% evaporation per hour [49].

## Aircraft Fueling

As noted earlier in this report and in line with airframe OEM reference materials [51], the target aircraft fueling time is 10 minutes. The fuel truck lines, aircraft fuel intake ports, and fuel tank supply lines are all sized to a nominal 2 in diameter. Once fuel is onboard the aircraft, additional splits in fuel distribution from each port are designed to allow the maximum flow from the fueling truck to reach all tanks or partitioned tank segments. Modern airline refueling practices use two hoses simultaneously to increase refueling rates; current analyses for the cases investigated later in this report find that the 10 minute refueling time is achievable without doubling hoses from each tanker truck, but this remains an option should larger, intercontinental $H_2$ aircraft be developed.

## Oxygen Supply: Atmospheric or Onboard

Most rockets have carried liquid oxygen onboard since the 1960s [52]. Supplying oxygen in exact quantities in the same manner as hydrogen allows for a more exact reaction, which could benefit any system. Further, it has been shown that oxy-combustion can result in higher efficiencies, since large amounts of nitrogen don't need to be heated and oxygen is not dilute [53]. However, the primary reason space vehicles carry oxygen is because they fly too high and fast to pull in oxygen from the atmosphere, problems that commercial aircraft do not face. Along with liquid oxygen supply comes the risk of system failures resulting in leaks of liquid $H_2$ and liquid oxygen, which can create dangerous explosive environments such as those tested by NASA in the wake of the space shuttle *Challenger* accident [52], [54].

For the reductions in complexity and risk, most studies of hydrogen aircraft presume that $O_2$ is supplied from the outside atmosphere [47]. This reduces the amount of gas to be stored onboard in liquid form, maximizing the amount of $H_2$ fuel to be carried and thereby maximizing the range of the aircraft while reducing the required increase in aircraft volume to accommodate the new fuel tanks. Using oxygen from the ambient air also reduces the energy used by the air transport system as a whole, eliminating the energy that would be needed to liquify $O_2$.

## Onboard Fuel Distribution

Every aircraft has a unique fuel system sized to its own requirements for fuel capacity and consumption, however all modern systems share many common elements to ensure safe, reliable operation. The $H_2$ fuel distribution systems require some innovative elements not traditionally included in kerosene fuel systems. Figure 16 shows the high-level layout for two fuel distribution systems explored in the later aircraft case studies.

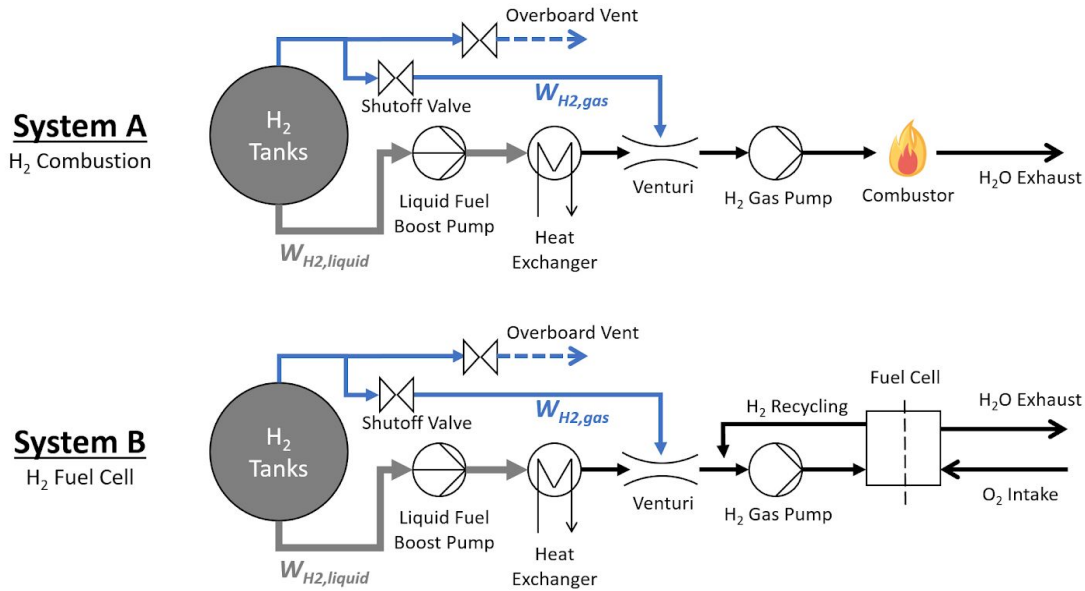

**Figure 16**: Fuel distribution systems for $H_2$ combustion (A) and $H_2$-driven fuel cells (B).

Both systems feature a mostly common layout, including tanks, boost pumps, heat exchangers, and a venturi section. Tanks are designed as described in the last section, capable of storing liquid $H_2$ with a boil off rate of about 0.05-0.08 %/hr (by mass). All commercial aircraft would feature multiple tanks or tank segments, with the exact quantity depending on range and payload requirements. Insulated supply lines carry fuel from the tanks to liquid fuel boost pumps, designed as a double-layer vacuum tube to minimize heat transfer to the flowing fuel. The boil off rate is well below the fuel consumption rates, so the $H_2$ gas escaping can be combined later in the fuel supply system (using a venturi to lower the main supply pressure and pull the excess $H_2$ in) so that it is not lost/wasted.

The boost pumps ensure a steady stream of liquid $H_2$ fuel leaves the tank to support the eventual fuel consumption rate. Liquid $H_2$ fuel is stored and initially pumped at 20.28 K. The boost pumps run on about 2-3 kW; most of this goes into raising the fuel pressure, with minimal temperature rise observed in the fuel, keeping the fuel in a liquid regime. Next the fuel passes to a heat exchanger (or series of heat exchangers) to raise the temperature of hydrogen and convert it into a gas. Once the fuel is out of storage, flowing fuel could enter a gaseous state as long as it continues flowing to the combustor to avoid dangerous pressure buildups. There are many potential sources of heat on the aircraft. One common and great candidate is the avionics; though impractical to route the $H_2$ fuel through the cockpit and then back to the engines, a secondary working fluid (such as compressed air or nitrogen) could circulate heat to the heat exchangers. After the venturi, an additional pump supplies the combustor with a regulated output pressure (and thereby regulated mass flow) of $H_2$ fuel to be burned in the combustor.

For turbine-propelled aircraft, the full supply of hydrogen fuel is used up (either burned or expelled if combustion is not perfectly stoichiometric). However for fuel-cell powered systems, some of the hydrogen gas is not used to processed into power and expelled as water, due to the saturation limitations of fuel cell operation. It would be possible to vent the unused $H_2$ gas overboard, but at a significant cost to SFC because the fuel vented must be considered as part of the consumption rate.

# Thrust Production using Hydrogen Fuel

## Combustion in a Turbine Engine

Traditional turbine engine combustors can run on hydrogen fuel, but the systems are optimized for kerosene jet fuels and therefore see a decrease in performance. This is attributed in large part to the differences in fuel-air ratio (FAR) for the kerosene and hydrogen combustion reactions, as detailed below.

Kerosene*: $\quad 2C_{12}H_{26}(l) + 37\{O_2(g) + 3.76N_2(g)\} \rightarrow 24CO_2(g) + 26H_2O(g) + 139.12N_2(g)$

Hydrogen: $\quad 2H_2(g) + \{O_2(g) + 3.76N_2(g)\} \rightarrow 2H_2O(g) + 3.76N_2(g)$

It should be noted that kerosene is represented here as dodecane. Kerosene combustion models usually include a mixture of gases, such as dodecane, iso-cetane, toluene, and others, with dodecane constituting the majority of the mixture [55]. Empirical data for aircraft engine fuel burn was used for all calculations in this paper, but the reaction of dodecane illustrates the vast differences in fuel molecule size, and therefore air content required, to run the turbine.

High-bypass turbofan engines used on modern commercial airliners produce thrust based largely on mass flow. The engines operate by running a relatively low amount of air through the core, to power a larger fan that produces 5-10 times as much thrust as the core. Engine cores burning $H_2$ fuel require only 77% of the mass flow to produce the same energy output driving the core, allowing for lower specific fuel burn rates (by mass) and the potential for greater bypass ratios in $H_2$ engines producing the same thrust as their kerosene-burning counterparts. To maintain overall engine thrust output, the core and fan flow must then be scaled up, resulting in final core mass flow at about 88% of an equivalent kerosene engine, with $H_2$ SFC (by mass) around 41% that of kerosene.

Further engine optimization would use a specialized combustor, designed to burn leaner than stoichiometric to keep temperatures down [47]. The lower temperature exhaust gases are compensated by the relative increase in specific heat of the combustion products (when compared to kerosene), largely due to the high amount of water vapor. The exhaust gas therefore sees a lower pressure drop across the turbine; despite turbine inlet temperatures about 60K lower than traditional turbine engines [48], hydrogen-burning turbines can produce more thrust for significant fuel savings. Research shows $H_2$ presents the potential to reduce specific fuel consumption (SFC) of nearly a factor of 3, when compared to a kerosene-burning engine producing the same thrust [56]. Hydrogen combustion also shows potential for about a 1% improvement on specific energy consumption (SEC), again when producing the same thrust [56]. These numbers don't quite hold when investigated further for specific cases, as will be explored in two flight case studies later in this paper, but the trend of better SFC (by weight) holds for all cases investigated.

A unique challenge of $H_2$ fuel is its potential to flash back, or spread combustion up the supply line. The risk is low during operation but presents more significant challenges at startup and shutdown, when supply lines from the cryogenic $H_2$ storage tanks could fill with ambient air. The design detailed previously would also include firewall shutoff valves at the tanks and a suction-feed fuel system to facilitate a safe fuel system design, however a successful and recommended method to further mitigate this risk involves flushing the supply lines with an inert gas such as nitrogen [57]. It is possible that this gas could be carried onboard the aircraft in small quantities, and this would likely be a required safety mechanism for widespread deployment of hydrogen-powered aircraft, so nitrogen storage and supply equipment should be factored into detailed aircraft designs.

## Fuel Cells and Hybrid-Electric Flight

Hydrogen aircraft could also be powered by fuel-cell driven electric motors. A hybrid fuel cell electric system has the advantage over combustion engines of not producing $NO_X$ at the lower reaction temperatures. The disadvantage is that it requires an additional step of power transfer, converting chemical energy first to electrical energy and then finally to mechanical energy. The value or lack thereof in this tradeoff depends on the specifications of available fuel cells and electric motors.

Typical fuel cells currently have specific power density of around 1.6 kW·kg$^{-1}$, though that is estimated to rise as high as 8 kW·kg$^{-1}$ in coming years [58]. The current world record specifications for a PEMFC stack (designed by Toyota) are 2.0 kW·kg$^{-1}$ and 3.1 kW·L$^{-1}$, though that was recently surpassed on a single-cell basis by a team at the Chinese Academy of Sciences who developed a small flexible cell capable of 2.23 kW·kg$^{-1}$ and 5.19 kW·L$^{-1}$ [59]. Comparing to a Boeing 737 with approximately 20,000L fuel capacity and conservatively assuming fuel cell installation in the wings would require a 25% knockdown for geometry, a 737 could hold capacity to generate 77.85MW - 22% over the a typical mission power requirement [60] - but this would weigh about 2.8x the weight of the previous jet fuel load, as kerosene fuel is one of the most energy dense fuel sources available.

Electric motors are also heavy equipment, but advances have brought the specific power up to 5.2 kW·kg$^{-1}$ with projected innovation expected to bring that up to 10 kW·kg$^{-1}$ in the future [58]. The viability of fuel cell power is assessed for the regional Embraer E175 later in this paper. Further design study and innovation in the enabling technologies is still required, but they are approaching a stage of maturity that may allow for development of a viable commercial hydrogen aircraft in the near future.

## Exhaust Products and Impact to Climate Change

In both $H_2$ combustion and fuel cell operation, the main reaction product is water. Combustion also yields a small amount of nitrogen oxides ($NO_X$), production of which is minimized through lean burning and lower flame temperatures [47]. However, the water (generally exhausted as water vapor) and $NO_X$ produced are not harmless. The Global Warming Potential (GWP) of water vapor, $NO_X$, and $CO_2$, on a 100-year basis, has been compared [47], [57]. Water vapor is harmless (GWP=0.00) below altitudes of about 10 km, but reaches GWP = 0.72 at 15 km. It may be possible to assess the performance of hydrogen-powered aircraft at lower-than-normal altitudes, to operate aircraft in ranges with $H_2O$ GWP=0.00. $NO_x$ is produced at very low quantities in well-designed systems, but carries GWP values on the order of 10-70 at most commercial aircraft operating altitudes.

## Heat Transfer to Liquid $H_2$ Prior To Use

Hydrogen molecules ($H_2$) can exist in two spin isomeric forms, Para $H_2$ and Ortho $H_2$. Para $H_2$ is the lower energy form of the two. At extremely low temperatures i.e. when stored in equilibrium as a liquid, most of the hydrogen is Para $H_2$, as shown in Figure 17. Heating the $H_2$ fuel before introducing it in the combustor of a turbofan engine was shown to save up to 12% fuel by mass and thereby increasing the maximum flight time for a given aircraft [61]. This was accomplished using heat exchangers to capture some of the excess exhaust heat, recouping between 1.2 - 3.2% of the LHV of hydrogen. Note that the specific heat of $H_2$ is high in all isomeric forms, as shown in Figure 17; additional heat sources could be explored to capture more heat before the fuel is burned.

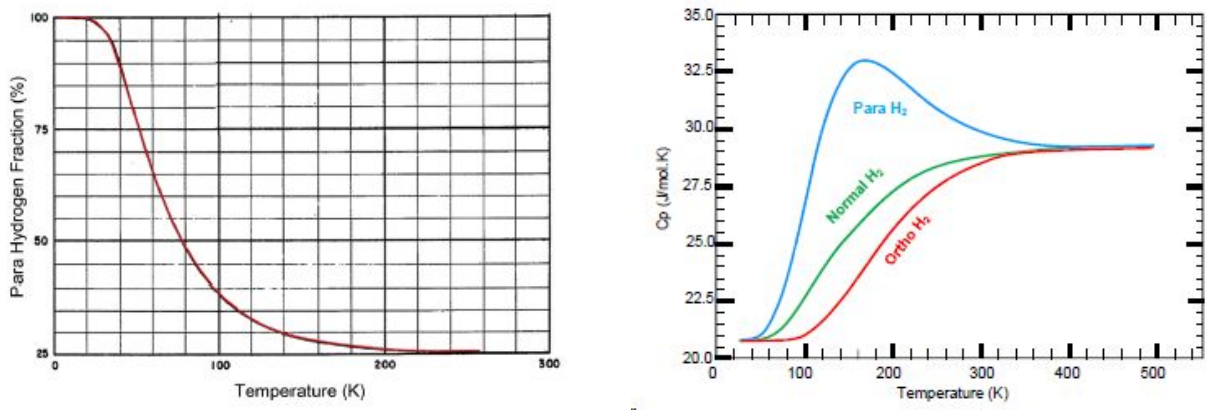

**Figure 17**: Mole fraction (left) and specific heat (right) of Hydrogen isomers. As temperature increases, more Ortho $H_2$ is present, yet specific heats are high for all forms [61].

Additional and unrelated innovations in air travel may also support the ability to heat hydrogen fuel without the need to use power generated by the engine or fuel cell. One such example is the ramjet engine, a simple engine design with no turbomachinery that only operates at supersonic speeds. Ramjet engines compress incoming air as it passes through shock waves and a convergent/divergent duct, before introducing fuel in the burner to add energy for thrust production [62]. Exergy study of one conceptual ramjet design found that the air temperature rose over 300K while slowing down through the inlet as shown in Figure 18, resulting in over 2MJ·s$^{-1}$ of heat generation, some of which could be dissipated into incoming hydrogen fuel ahead of the combustion chamber should there be a design advantage to doing so. This concept is not explored in this study, but given the recent interest in developing supersonic civil aircraft [63] [64] [65] and the potential synergies of supersonic inlet heat and hydrogen fuel, supersonic $H_2$-fueled designs should be explored further.

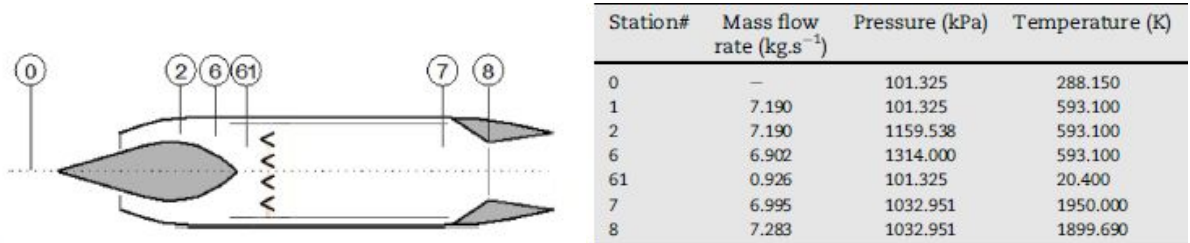

**Figure 18**: Station diagram of a ramjet engine, with cycle data [66].

# Case Studies: Two Common Routes from LAX

Two flights were assessed for the LAX case study: a relatively short flight between Los Angeles and Las Vegas, and a relatively long flight from Los Angeles to New York City (JFK). Both flights are high-volume routes that have multiple scheduled flights on multiple carriers daily, with easy access to historical data. As shown in Figure 19, these routes are also entirely overland, so they don't require ETPOS (Extended-range Twin-engine Operational Performance Standards) planning and the historical safety records that allow for these operations as designated in by FAA regulation (14 CFR part 121.193). Both are great candidates for an introduction of hydrogen-powered flight.

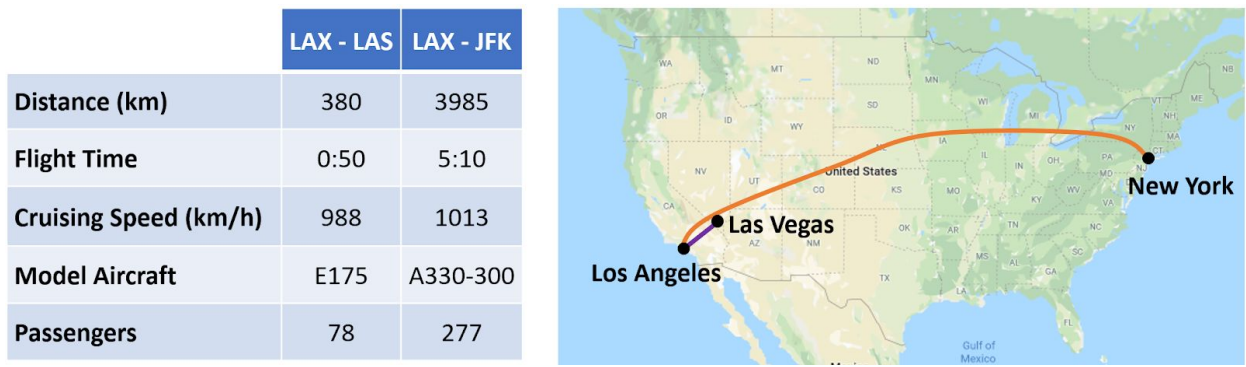

**Figure 19**: Common regional and transcontinental routes from LAX.

## Case 1: Regional Flight (LAX to LAS)

The flight from Los Angeles to Las Vegas is a frequently traveled route, with over 30 daily scheduled flights on multiple carriers. The route is a relatively-straight shot of about 380 km (236 mi), not including unexpected air traffic control routings. With a flight time of only 50 minutes, aircraft do not reach the high altitudes achieved by longer flights but instead top out around 7620 m (25,000 ft). In addition, very little of the trip is spent with the aircraft in cruise, when it is most efficient; aircraft making a typical trip spend 15 minutes climbing, only 10 minutes cruising at altitude, then descend for about 25 minutes before landing. Aircraft making this trip are all narrow-body aircraft, typically seating 75 to 175 passengers. This analysis uses the Embraer E175 as the base aircraft, a 78 passenger jet traditionally powered by two GE CF34-8E high-bypass turbofan engines.

## Fuel Consumption and Engine Design

Running on hydrogen fuel decreases the burn rate (by mass) required in all segments of flight. A typical E175 runs on two CF34-8E turbofan engines, rated to 14,500 lbf (64.5 kN) of thrust. Based on the mission requirements and time segments, the flight typically requires around 2162 kg (2689 L) of Jet A-1 fuel. FAA regulations (14 CFR 91.167) also require passenger aircraft to carry fuel reserves that enable the aircraft to divert to a backup destination, and cruise for 45 minutes after reaching the backup airport. Required reserves constitute a significant additional fuel requirement on shorter flights: with adequate reserves, a traditional E175 must carry 4,454 kg (5,540 L), over twice the amount needed for the trip itself.

Using similar engines optimized for hydrogen combustion, the fuel requirements (by weight) decrease. The CF34-8E has a 5:1 bypass ratio, but optimizing for $H_2$ fuel increases this to a new 5.75:1 bypass ratio. The hydrogen engines have a resulting SFC that is 41.2% of the kerosene SFC. Resulting fuel requirements for the trip are 1,835 kg (13,491 L) including required reserves - meaning a liquid hydrogen regional jetliner needs about 2.5 times the fuel capacity (volume) compared to its kerosene-burning counterpart.

## Aircraft Design

The fuel volume storage requirements drive changes to the aircraft design. The E175 fuselage is optimized for regional transport today, meaning the passenger cabin takes up most of the fuselage with relatively-low cargo space under the floor. As is typical of modern jetliners, the E175 features a "wet wing" design where the kerosene fuel is stored inside the wings. This design is an efficient use of space, but it incompatible with liquid hydrogen fuel due to its low density and storage temperature. The wings would not be able to fit all of the fuel, so additional fuselage-held tanks would be required. More importantly, the flexing action of the wings in flight would limit the materials available for tank insulation, leading to higher heat flux and $H_2$ boil-off rates, as well as elevated risk of ice formation on the wings.

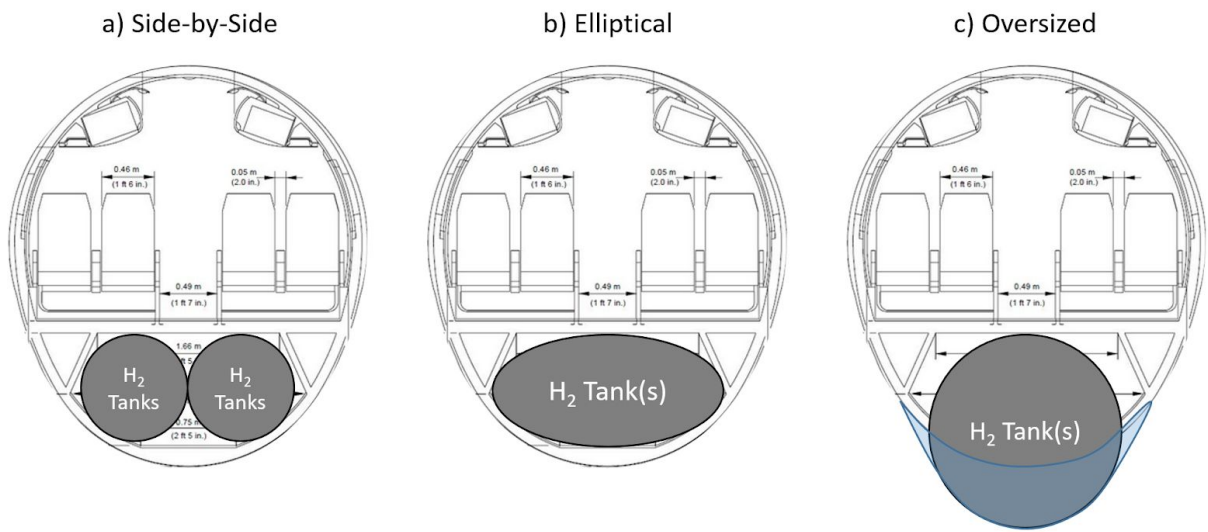

**Figure 20**: Potential $H_2$ tank layouts include a) side-by-side, b) elliptical, and c) oversized tanks.

Fuel storage inside the fuselage is a better option for liquid $H_2$. As shown in Figure 20, there are multiple potential configurations with varying impacts on the aircraft design. Options (a) and (b) would fit within the current fuselage cross section with no impact to passenger capacity, however they would take up the majority of available below-deck storage. Design (a) would require 71 spherical tanks, each 0.94 m in diameter, or two cylindrical tanks with hemispherical end caps each 21.7 m in length, roughly the maximum length the fuselage could accommodate without a change in the outer profile. The tank length could be slightly decreased using an elliptical profile that more effectively fills the fuselage space as in design (b), but the gains are not enough to make the design practical.

Option (c) requires an 12% increase in frontal area as visualized in Figure 20, so a 4% increase in fuel burn is assumed required to compensate for the increase in parasitic drag for all phases of flight. This is likely compensated at least partially by decreased induced drag (from lower overall weight requiring lower lift for flight), but this is conservatively not accounted for. The resulting fuel tank holding enough $H_2$ for the flight with reserves would be 14.0 m long, and remove about 50% of the below-deck cargo area. On regional flights many passengers are on short trips and bring only carry-on sized luggage, so this decrease may be acceptable for airline operations; this should be investigated further.

If the tank length increased to 20.5 m, a length still compatible with the fuselage length but eliminating all below-deck cargo space, the aircraft could hold enough $H_2$ for a round trip flight with reserves at the end. This would allow the aircraft to refuel only at LAX and legally operate the route on an out-and-back basis. Many carriers do not fly regional aircraft in this manner, but instead fly the aircraft on a long route with multiple stops along the way. However, if the airline set up one aircraft to fly its currently-scheduled multiple flights between LAX and LAS on an out-and-back basis without allowing checked luggage, the aircraft could accomplish that flight on a regular basis. Not allowing carry-on luggage would not work for all passengers; it may be possible to fly half of the daily flights using this aircraft, using a conventional aircraft for the other flights to accommodate passengers with more luggage.

This largest tank has an increased boil off rate compared to the original smaller tanks described in earlier sections [49], but it is still easily manageable. The 20.5 m tank would see boil off rates of about 2.4 g/s based on increased heat transfer through the larger tank surface area and additional fuel pipes. This is just 3.2% of the fuel burn rate at idle, and 0.5% of the burn rate during takeoff and climb. The fuel system can incorporate this boil rate into the fuel supply, resulting in no loss of usable fuel to boil off in flight.

The changes in engine design would have minimal effect on overall aircraft design. While the bypass ratio increases, the absolute size of the engine core (which is the densest/heaviest part of the engine) descreases. Overall net effect on weight is assumed to be negligible. The increase in engine fan size may result in a need for taller landing gear, but this has already been accounted for by the increased height needed to prevent the current gear size should suffice.

All together, the fuel tanks are the main driver of increased aircraft cost. Airframe materials and wing designs would be largely unchanged, unfortunately losing the economic option of using the wings as fuel tanks. Fuel hydraulic systems, though modified for $H_2$, would include roughly the same amount of routing lines and powered components. The tanks would likely result in a 5-10% increase in aircraft CapEx. Based on current airline financing structure, most of the operating costs go to fuel, salaries/training, and airport gate rentals, with relatively little going to the cost of the physical aircraft in comparison. It is far more likely that safety and public perception of $H_2$ fuel would be larger barriers to industry and consumer acceptance of $H_2$ fueled aircraft.

## Fuel Cell with Electric-Driven Props

Due to the shorter distances covered and lower amount of power required to fly, regional jets are a potential application for fuel cell power instead of hydrogen combustion. Fuel cells raise the possibility of higher efficiency in power generation from $H_2$ (as compared with turbine combustion). Based on current technology, fuel cell density (weight) and inefficiencies in power transfer do not make fuel cells a better option than turbine engines.

Fuel cells can theoretically generate adequate power required to fly the aircraft, and the systems are fairly efficient. Typical fuel cells today can achieve around 60% efficiency [67], motor controllers are around 95% efficiency [67], [68], and brushless DC motors are around 99% efficiency. Once the motor generates torque, the propellers convert mechanical energy into thrust with efficiencies of about 81% [69]. In theory, ducted fans can be more efficient than unshrouded impellers because the shroud prevents the formation of vortices coming off of the propeller tips, however the added complexity and weight of a shrouded fan often negates some of the benefits. Currently there is significant research investigating ducted fan systems designed for urban airborne mobility vehicles, and the technology shows significant potential for increases in efficiency and decreases in fuel burn [70]; based on today's available technology, these potential gains were not included in this study. Therefore, overall efficiency of a fuel cell system producing thrust for an aircraft comes to roughly 45.7%, significantly better than the roughly 40% efficiency of turbine engines.

The resulting system requires about 88% of the fuel required by an aircraft powered by $H_2$ combustion turbines, a welcome reduction in fuel burn and required fuel storage volume. The issue is the additional weight of the fuel cells. With experimental fuel cell densities potentially as high as only 2.2 kW/kg [59], the fuel cell system would require about 19,800 kg of equipment to generate the 43.7 MW of power output during the climb phase of the flight, an impractical weight addition compared to the 2364 kg total for two CF34-8E turbine engines on the aircraft today. Lower-power climbs and denser fuel cells, or potentially a hybrid system using fuel cells for the base load and combustion for additional thrust during climb only, would be required to make this a practical solution.

# Case 2: Transcontinental Flight (LAX to JFK)

Flights from Los Angeles to New York City (JFK) are similarly busy, if not busier than the LAX-LAS route especially considering multiple airports connected directly between the LA and NYC areas. As seen in Figure 19, the typical route flies northwest to Idaho before turning east over Chicago and Detroit, then turning slightly southeast over New York until reaching JFK, to take advantage of the shorter distances covered by a great-circle routing. This 5 hour and 10 minute flight includes 25 minutes of climb, roughly four hours of cruise, and a 40 minute descent. The route is flown by both narrow- and wide-body aircraft, depending on the carrier and flight demand; this analysis is modeled after the Airbus A330-300, a 277 passenger aircraft powered by either two Rolls-Royce Trent 700 engines or two Trent 7000 engines on the new A330neo (A330-800).

## Fuel Consumption and Engine Design

Though this aircraft is larger than the Embraer E175 used in shorter regional flights, hydrogen fuel similarly decreases the burn rate (by mass) required in all segments of flight. The Trent 700 engine is a 5:1 bypass ratio, and would see a specific fuel consumption (SFC, by mass) of only 6.59 g/kN/s, only 41% that of the 16.0 g/kN/s the kerosene-burning engine is rated for. The newer, more efficient Trent 7000 has a 10:1 bypass ratio and, when adjusted for hydrogen combustion, would have SFC (by mass) of only 5.90 g/kN/s, a 10.5% improvement over the older Trent 700. The state-of-the-art Trent 7000 engine (and it's theoretical hydrogen-burning version) are used for this analysis.

An A330-800 with RR Trent 7000 engines would require 49,979 kg (62,163 L) of Jet A-1 to make the LAX-JFK trip with adequate reserves. It is notable here that the FAA requirements for fuel reserves do not change with trip length; thus, the reserves constitute a much smaller 17% of the total fuel required for the transcontinental trip. With similar engines burning $H_2$ fuel (the redesigned engines would have an 11.4:1 bypass ratio), the aircraft would require 20,483 kg (292,618 L) of $H_2$ fuel.

## Aircraft Designs

If $H_2$ fuel tanks were designed to fit in the lower section of the fuselage, they would be 42.7 m in length, taking up the majority of below-deck cargo space similarly to the E175 $H_2$ tank design. This tank arrangement, as proposed for the regional jet, is not as practical for longer flights. Passengers making longer trips often bring more luggage with them, necessitating the use of below-deck cargo space. More importantly, air carriers often transport cargo for shipping companies to maximize aircraft capacity and earn more on the longer, expensive flights.

A better arrangement for the A330-800 would be to install two large fuel tanks, one near the front and one at the rear of the aircraft. Two tanks are important, to ensure the aircraft remains aerodynamically balanced fore-to-aft as the fuel is consumed and fuel weight in the tanks decreases. As shown in Figure 21, tanks of equal volume (146.3 m$^3$ each) in the forward section of the aircraft and the rear cone area would allow for this balance. This would also maintain the current profile of the aircraft with no need to increase fuselage size, and therefore no increase to drag and thrust as seen with the E175.

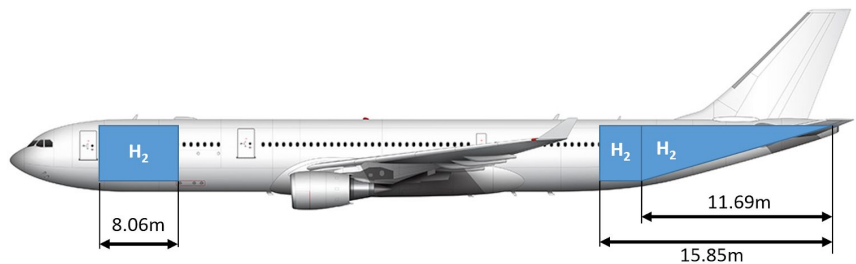

**Figure 21** : $H_2$ tank layout for A330 aircraft.

One difficulty with this tank arrangement is that the aircraft seating arrangement is compromised by the inclusion of tanks in the full fuselage section. The aircraft's normal seating arrangement includes a large area of first-class seating. For overnight ("red eye") flights, it would not be economical to remove these seats; for daytime flights, however, airlines could fit the full 277 passengers with a new section of "Eco-class" seats, as shown in Figure 22. Branded as a superior class to coach that enables the $H_2$ fueled aircraft to fly cleanly across the country, economic incentives could be set up for companies that book tickets in this section rather than the traditional business class seats.

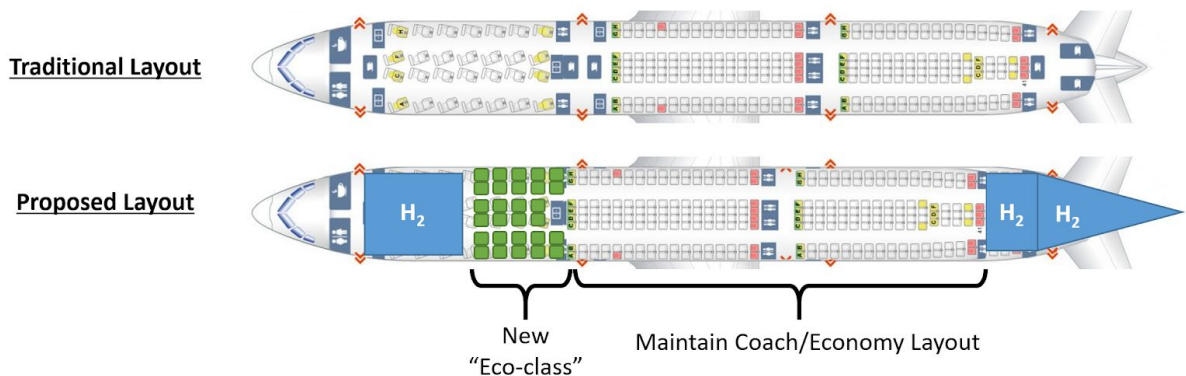

**Figure 22**: Updated seating layout, including "Eco-class" to maintain passenger capacity.

Currently the A330-300 requires 25 minutes for fuel pumping time including connecting/disconnecting the fuel lines (which are assumed to be at optimal time already). This is well above the 10 minutes previously assumed as part of the LAX fuel distribution, so the A330-300 aircraft running on liquid $H_2$ will require three fuel trucks to refuel. Multiple fuel inlets on the aircraft exterior are typical, and would not affect the boil off rate of the fuel. Simultaneous deployment of three fueling trucks would actually improve refueling time, and could thereby decrease aircraft turnaround time, over current aircraft.

Currently the A330-300 requires 25 minutes for fuel pumping time [51], [71] including connecting/disconnecting the fuel lines (which are assumed to be at optimal time already). This is well above the 10 minutes previously assumed as part of the LAX fuel distribution, so the A330-300 aircraft running on liquid $H_2$ will require three fuel trucks to refuel. Simultaneous deployment would actually improve refueling time, and could thereby decrease aircraft turnaround time, over current aircraft.

# Conclusion

Overall, we have shown that a conservative estimate of cost and performance of the infrastructure needed for hydrogen aircraft based on commercially deployed technology results in ~2X increase in fuel cost compared to jet fuel. Dominating costs include electricity and the CapEx of electrolysis and liquefaction. Assuming this fuel could enter production, it is feasible to build and operate aircraft that run on $H_2$, with the design changes necessitated by the switch, for both short- and long-range flights. On the other hand, reasonable and predicted improvements (or even deployments/economies of scale of existing technology) along with more efficient electricity regulation can change the result entirely, where we predict by 2030 that there could be a 25% *decrease* in fuel costs by switching from jet fuel to hydrogen. Further realistic technological advances, along with the potential for new government incentives to support hydrogen-fueled aircraft, could make this a legitimate possibility in the near future.

# Appendix A: LAX Hydrogen Supply, Base Case

| Base Case: Grid Electricity | | | | |
|---|---|---|---|---|
| | Variable | Value | Unit | Comment |
| Plant Sizing | LAX Fuel Use | 1.50E+09 | gal/yr | |
| | HHV Jet Fuel | 45 | MJ/kg | |
| | Jet Fuel density | 3.00 | kg/gal | |
| | LAX Fuel Use | 6411.68 | MW | Average Jet Fuel Energy use for Aviation at LAX |
| | HHV H2 | 142 | MJ/kg | |
| | H2 rate | 45.153 | kg/s | H2 reqd. assuming efficiency H2=Jet fuel |
| | | | | |
| Electricity | Energy Price | 0.0351 | $/kWh | 2019 Off peak price for large 50kV customer |
| | Delivery | 0.0175 | $/kWh | Trans, dist, fees - no demand charge by interuptable program |
| | Total | 0.0526 | $/kWh | |
| | Capacity Factor | 0.792 | | No use on-peak 4-9pm |
| | | | | |
| Electrolysis | Efficiency | 80% | | Assuming current AEC Electrolysis |
| | Energy Reqd. | 177.5 | MJ/kg | |
| | Power | 8014.6 | MW | |
| | Capital Cost | 1200 | $/kWel | Typical cost of PEM Electrolyzer, and 1 cell replacement, disc. |
| | Capital Cost | 1.21E+10 | $ | for LAX, before financing, considering capacity factor |
| | Life | 60000 | hours | |
| | Life | 18 | years | Lifetime, considering capacity factor, and 1 cell replacement |
| | Real Disc. Rate | 8% | | Assuming 10% discount rate, 2% inflation for investment ROI |
| | Financing Factor | 1.89 | | Effective Capital cost multiplier as a result of ROI |
| | Real capital cost | 2.29E+10 | $ | for LAX |
| | Electricity cost | 6.65E+10 | $ | assuming electricity price increases with inflation |
| | Total Cost | 8.94E+10 | $ | |
| | | | | |
| Liquefaction | Efficiency | 30% | | Linde Cycle --> improve with Claude or other |
| | Thermo. Min work | 12 | MJ/kg | Based on the change in exergy of the H2 |
| | Work reqd. | 40.0 | MJ/kg | |
| | Power | 1806.1 | MW | |
| | Capital cost | 3.94E+09 | $ | Based on NREL analysis for Linde Cycle, cons. cap factor |
| | Real capital cost | 7.44E+09 | $ | Assuming same life as electrolysis plant |
| | Electricity cost | 1.50E+10 | $ | assuming electricity price increases with inflation |
| | Total Cost | 2.24E+10 | $ | |

| Storage | Duration | 168 | hours | 1 week, buffer prices/supply/demand/maintenance |
|---|---|---|---|---|
| | Size | 2.73E+07 | kg | Stored at generation site, at LAX |
| | Density | 70 | kg/m3 | at 4 bar saturated liquid |
| | Volume | 390119 | m3 | |
| | Diameter | 20.0 | m3 | To match the current largest tanks, e.g. at NASA |
| | Number of Tanks | 93 | | This is a ~250x250m footprint, can easily be located onsite |
| | Surface Area | 1.17E+05 | m2 | |
| | Heat transfer | 2.64E-02 | MW | Based on vacuum conditions and 10 97% reflective shields |
| | H2 latent heat | 0.461 | MJ/kg | |
| | Boil rate | 5.72E-02 | kg/s | |
| | % Loss | 0.13% | | could be piped back to Liquefaction plant, but loss is very low |
| | Capital Cost | 8.65E+08 | $ | Based on NREL cost of similar sized tank, inflation adj to 2019 |
| | Real capital cost | 1.63E+09 | $ | |
| | | | | |
| Transport | H2 rate | 162550 | kg/hr | Transported 0-3 miles to aircraft, assume 2 trips per hour |
| | Truck Capacity | 4300 | kg | current semi-truck sized liquid hydrogren vehicle |
| | Min # of trucks | 19 | | Number of trucks to distribute fuel, if planes left sequentially |
| | Max # of trucks | 128 | | 2 trucks needed planes (737 needs 1.5 trucks), 128 gates |
| | Typ. # of trucks | 100 | | Reasonable compromise |
| | Truck Cost | 167 | $/kg | assuming the trucks last as long as the plant |
| | Truck Cost | 718100 | | |
| | Capital Cost | 7.18E+07 | $ | |
| | Real capital cost | 1.36E+08 | | |
| | Driver Cost | 1.80E+08 | | 2 full time drivers per truck, wages+ inflation, $50K/driver |
| | Total Cost | 3.16E+08 | | |
| | | | | |
| Totals | Total CAPEX | 3.21E+10 | $ | $30B |
| | Lifetime H2 Produced | 2.56E+10 | kg | |
| | Electrolysis CapEx | 0.89 | $/kg | |
| | Electrolysis Electricity | 2.60 | $/kg | |
| | Liquefaction CapEx | 0.29 | $/kg | |
| | Liquefaction Electricity | 0.58 | $/kg | |
| | Storage Cost | 0.06 | $/kg | |
| | Transportation Cost | 0.01 | $/kg | |
| | Total Cost | 4.44 | $/kg | |
| | Total Cost | 31.27 | $/GJ | |
| | Cost Jet Fuel | 2 | $/gal | Based on $2/gal |
| | Cost Jet Fuel | 14.8 | $/GJ | |
| | H2/Jet fuel | 2.11 | | Greater than 1 indicated Hydrogen is more expensive |

# Appendix B: LAX Hydrogen Supply, 2030 Case

| Base Case: Grid Electricity | | | | |
|---|---|---|---|---|
| | Variable | Value | Unit | Comment |
| Plant Sizing | LAX Fuel Use | 1.50E+09 | gal/yr | |
| | HHV Jet Fuel | 45 | MJ/kg | |
| | Jet Fuel density | 3.00 | kg/gal | |
| | LAX Fuel Use | 6411.68 | MW | Average Jet Fuel Energy use for Aviation at LAX |
| | HHV H2 | 142 | MJ/kg | |
| | H2 rate | 45.153 | kg/s | H2 reqd. assuming efficiency H2=Jet fuel |
| | | | | |
| Electricity | Energy Price | 0.0150 | $/kWh | 2019 Off peak price for large 50kV customer |
| | Delivery | 0.0088 | $/kWh | Assume on site generation (mostly offshore wind) |
| | Total | 0.0238 | $/kWh | |
| | Capacity Factor | 0.792 | | No use on-peak 4-9pm |
| | | | | |
| Electrolysis | Efficiency | 85% | | Assuming future AEC/PEMEC Electrolysis |
| | Energy Reqd. | 167.0588235 | MJ/kg | |
| | Power | 7543.1 | MW | |
| | Capital Cost | 500 | $/kWel | Typical cost of PEM Electrolyzer, and 1 cell replacement, disc. |
| | Capital Cost | 4.76E+09 | $ | for LAX, before financing, considering capacity factor |
| | Life | 90000 | hours | |
| | Life | 26 | years | Lifetime, considering capacity factor, and 1 cell replacement |
| | Real Disc. Rate | 1% | | Assuming 10% discount rate, 2% inflation for investment ROI |
| | Financing Factor | 1.14 | | Effective Capital cost multiplier as a result of ROI |
| | Real capital cost | 5.41E+09 | $ | for LAX |
| | Electricity cost | 4.09E+10 | $ | assuming electricity price increases with inflation |
| | Total Cost | 4.63E+10 | $ | |
| | | | | |
| Liquefaction | Efficiency | 60% | | Claude cycle |
| | Thermo. Min work | 12 | MJ/kg | Based on the change in exergy of the H2 |
| | Work reqd. | 20.0 | MJ/kg | |
| | Power | 903.1 | MW | |
| | Capital cost | 2.46E+09 | $ | Extrapolated Shimko design, considering cap factor |
| | Real capital cost | 2.80E+09 | $ | Assuming same life as electrolysis plant |
| | Electricity cost | 4.89E+09 | $ | assuming electricity price increases with inflation |
| | Total Cost | 7.69E+09 | $ | |
| **Storage** | Duration | 24 | hours | 1 day, buffer prices/supply/demand |
| | Size | 3.90E+06 | kg | Stored at generation site, at LAX |
| | Density | 70 | kg/m3 | at 4 bar saturated liquid |
| | Volume | 55731 | m3 | |
| | Diameter | 20.0 | m3 | To match the current largest tanks, e.g. at NASA |
| | Number of Tanks | 13 | | This is a ~250x250m footprint, can easily be located onsite |
| | Surface Area | 1.63E+04 | m2 | |
| | Heat transfer | 3.69E-03 | MW | Based on vacuum conditions and 10 97% reflective shields |
| | H2 latent heat | 0.461 | MJ/kg | |
| | Boil rate | 7.99E-03 | kg/s | |
| | % Loss | 0.02% | | could be piped back to Liquefaction plant, but loss is very low |
| | Capital Cost | 1.24E+08 | $ | Based on NREL cost of similar sized tank, inflation adj to 2019 |
| | Real capital cost | 1.40E+08 | $ | |
| | | | | |
| **Transport** | H2 rate | 162550 | kg/hr | Transported 0-3 miles to aircraft, assume 2 trips per hour |
| | Truck Capacity | 4300 | kg | current semi-truck sized liquid hydrogren vehicle |
| | Min # of trucks | 19 | | Number of trucks to distribute fuel, if planes left sequentially |
| | Max # of trucks | 128 | | 2 trucks needed planes (737 needs 1.5 trucks), 128 gates |
| | Typ. # of trucks | 100 | | Reasonable compromise |
| | Truck Cost | 83.5 | $/kg | assuming the trucks last as long as the plant |
| | Truck Cost | 359050 | | |
| | Capital Cost | 3.59E+07 | $ | |
| | Real capital cost | 4.08E+07 | | |
| | Driver Cost | 2.60E+08 | | 2 full time drivers per truck, wages+ inflation, $50K/driver |
| | Total Cost | 3.01E+08 | | |
| | | | | |
| **Totals** | Total CAPEX | 8.39E+09 | $ | $30B |
| | Lifetime H2 Produced | 3.70E+10 | kg | |
| | Electrolysis CAPEX | 0.15 | $/kg | |
| | Electrolysis Electricity | 1.10 | $/kg | |
| | Liquefaction CAPEX | 0.08 | $/kg | |
| | Liquefaction Electricity | 0.13 | $/kg | |
| | Storage Cost | 0.00 | $/kg | |
| | Transportation Cost | 0.01 | $/kg | |
| | Total Cost | 1.47 | $/kg | |
| | Total Cost | 10.35 | $/GJ | |
| | Cost Jet Fuel | 2 | $/gal | Based on $2/gal |
| | Cost Jet Fuel | 14.8 | $/GJ | |
| | H2/Jet fuel | 0.70 | | Greater than 1 indicated Hydrogen is more expensive |

# Appendix C: EES Model of Claude Cycle

```
1:  "Constants:"
2:  eta_c = .8
3:  eta_t = .9
4:  R = 100
5:  y=0.5
6:
7:  "State 1"
8:  T[1] = 300 [K]
9:  P[1] = 101.3 [kPa]
10: h[1] = enthalpy(Hydrogen,T=T[1],P=P[1])
11: s[1] = entropy(Hydrogen,T=T[1],P=P[1])
12:
13: "State 2"
14: T[2] = T[1] "isothermal compressor"
15: P[2] = P[1]*R
16: h[2] = enthalpy(Hydrogen,T=T[2],P=P[2])
17: s[2] = entropy(Hydrogen,T=T[2],P=P[2])
18:
19: w_c = (h[1]-h[2] -T[1]*(s[2]-s[1]))/eta_c
20:
21: "State 3"
22: P[3] =P[2]
23: T[3] = 160 [K]
24: T[3] = temperature(Hydrogen,h=h[3],P=P[3])
25: s[3] = entropy(Hydrogen,T=T[3],P=P[3])
26:
27: "State 4"
28: P[4] = P[3]
29: T[4] = temperature(Hydrogen,h=h[4],P=P[4])
30: s[4] = entropy(Hydrogen,T=T[4],P=P[4])
31:
32: "State 5"
33: P[5] = P[4]
34: T[5] = temperature(Hydrogen,h=h[5],P=P[5])
35: s[5] = entropy(Hydrogen,T=T[5],P=P[5])
36:
37: "State 6"
38: P[6] = P[1]
39: T[6] =t_sat(Hydrogen,P=P[6])
40: h[6] = h[5]
41: h[6] = enthalpy(Hydrogen,P=P[6],x=x)
42: s[6] = entropy(Hydrogen,P=P[6],x=x)
43:
44: "State 7"
45: T[7] = T[6]
46: P[7] = P[6]
47: h[7] = enthalpy(Hydrogen,P=P[7],x=0)
48: s[7] = entropy(Hydrogen,P=P[7],x=0)
50: "State 8"
51: T[8] = T[6]
52: P[8] = P[6]
53: h[8] = enthalpy(Hydrogen,P=P[8],x=1)
54: s[8] = entropy(Hydrogen,P=P[8],x=1)
55:
56: "State 9"
57: P[9] = P[8]
58: T[9] = temperature(Hydrogen,h=h[9],P=P[9])
59: s[9] = entropy(Hydrogen,T=T[9],P=P[9])
60:
61: "State 10"
62: P[10] =P[9]
63: h[10] = enthalpy(Hydrogen,T=T[10],P=P[10])
64: s[10] = entropy(Hydrogen,T=T[10],P=P[10])
65:
66: "State 11"
67: T[11] =T[1]
68: P[11] = P[1]
69: h[11] = h[1]
70: s[11] =s[1]
71:
72:
73: "HXs"
74: h[2]-h[3] = (y*x+(1-y))*(h[11]-h[10])
75: y*(h[3]-h[4]) = (y*x+(1-y))*(h[10]-h[9])
76: y*(h[4]-h[5]) = y*x*(h[9]-h[8])
77: h[6] = (1-x)*h[7] + x*h[8]
78:
79: "Turbine"
80: s9s = s[3]
81: h9s = enthalpy(Hydrogen,P=P[9],s=s9s)
82: w_t_s = h[3]-h9s
83: w_t = eta_t*w_t_s
84: h[9] = h[3] -w_t
85:
86: Energy = (w_c - (1-y)*w_t)/(y*(1-x))
87: Exergy = h[7]-h[1] -T[1]*(s[7]-s[1])
88:
89: Efficiency =Exergy/Energy
```